\documentclass[reprint, aps, prb, footinbib, letterpaper, superscriptaddress, floatfix]{revtex4-2}

\usepackage[T1]{fontenc} 

\usepackage{amsmath,color, tikz}
\usetikzlibrary{matrix}
\usepackage{amssymb}
\usepackage{amsfonts}
\usepackage{amsthm}
\usepackage{mathtools}
\usepackage{comment}
\usepackage{hyperref}
\usepackage{cleveref}

\usepackage{pythonhighlight}

\usepackage{algorithm}
\usepackage{algpseudocode}
\usepackage{dsfont}

\usepackage{bbold}
\usepackage{chemformula}
\usepackage{siunitx}

\DeclarePairedDelimiterX\braket[2]{\langle}{\rangle}{#1 \delimsize\vert #2}
\DeclarePairedDelimiterX\braket3[3]{\langle}{\rangle}{#1 \delimsize\vert #2 \delimsize\vert #3}

\usepackage{accents}
\usepackage{xspace}

\usepackage{fancyvrb}

\usepackage{pdfpages}

\newcommand{\vJ}{\mathbf{J}}
\newcommand{\vP}{{\boldsymbol{\mathcal{P}}}}

\newcommand{\vE}{\mathbf{E}}

\newcommand{\vB}{\mathbf{B}}

\newcommand{\vD}{\mathbf{D}}
\newcommand{\ve}{\mathbf{e}}

\newcommand{\vnabla}{\boldsymbol{\nabla}}

\newcommand{\vmu}{\boldsymbol{\mu}}

\newcommand{\vkappa}{\boldsymbol{\kappa}}
\newcommand{\vH}{\mathbf{H}}

\newcommand{\vR}{\mathbf{r}}

\newcommand{\vk}{\mathbf{k}}

\newcommand{\vxi}{\boldsymbol{\xi}}

\newcommand{\hH}{\hat{H}}

\newcommand{\tr}[1]{\text{Tr}\left(#1\right)}

\newcommand{\avg}[1]{\left\langle #1\right\rangle}

\newcommand{\MaxwellLink}{\textsc{MaxwellLink}\xspace}

\usepackage{listings}
\usepackage{xcolor}

\usepackage{makecell}       

\definecolor{codegreen}{rgb}{0,0.6,0}
\definecolor{codegray}{rgb}{0.5,0.5,0.5}
\definecolor{codepurple}{rgb}{0.58,0,0.82}
\definecolor{backcolour}{rgb}{0.95,0.95,0.92}

\lstdefinestyle{mystyle}{
    backgroundcolor=\color{backcolour},   
    commentstyle=\color{codegreen},
    keywordstyle=\color{magenta},
    numberstyle=\tiny\color{codegray},
    stringstyle=\color{codepurple},
    basicstyle=\ttfamily\footnotesize,
    breakatwhitespace=false,         
    breaklines=true,                 
    captionpos=b,                    
    keepspaces=true,                 
    showspaces=false,                
    showstringspaces=false,
    showtabs=false,                  
    tabsize=2
}

\makeatletter
\AtBeginDocument{\let\LS@rot\@undefined}
\makeatother

\begin{document}

    \title{MaxwellLink: A Unified Framework for Self-Consistent Light-Matter Simulations}

\author{Xinwei Ji}
        \thanks{These authors contributed equally to this work.}
        \affiliation{Department of Physics and Astronomy, University of Delaware, Newark, Delaware 19716, USA}

    \author{Andres Felipe Bocanegra Vargas}
    \thanks{These authors contributed equally to this work.}
        \affiliation{Department of Physics and Astronomy, University of Delaware, Newark, Delaware 19716, USA}

        \author{Gang Meng}
        \affiliation{Department of Physics and Astronomy, University of Delaware, Newark, Delaware 19716, USA}
        
	   \author{Tao E. Li}%
	   \email{taoeli@udel.edu}
	   \affiliation{Department of Physics and Astronomy, University of Delaware, Newark, Delaware 19716, USA}
    
    \begin{abstract}
         A major challenge in light-matter simulations is  bridging the disparate time and length scales of  electrodynamics and  molecular dynamics. Current computational approaches often rely on  heuristic approximations of either the electromagnetic (EM) or the material component, hindering the exploration of complex light-matter systems. Herein,  \MaxwellLink \ --- a modular, open-source Python framework --- is developed for the massively parallel, self-consistent propagation of classical EM fields interacting with a large heterogeneous molecular ensemble. The package utilizes a robust TCP/UNIX socket interface to couple EM solvers  with a wide range of molecular drivers.  In this initial release, \MaxwellLink supports EM solvers spanning from single-mode cavities to full-feature three-dimensional finite-difference time-domain (FDTD) engines, and molecules described by multilevel open quantum systems, force-field and first-principles molecular dynamics, and nonadiabatic real-time Ehrenfest dynamics.  With the socket-based  architecture, users can seamlessly switch between levels of theory of either the EM solver or molecules without modifying the counterpart.  Moreover,  the EM engine and molecular drivers scale independently across multiple high-performance computing (HPC) nodes, facilitating large-scale simulations previously inaccessible to existing numerical schemes. The versatility and accuracy of this code are further demonstrated through applications including superradiance, radiative energy transfer, vibrational strong coupling in Bragg resonators, and plasmonic heating of molecular gases.  By providing a unified, extensible engine, \MaxwellLink potentially offers a powerful platform for exploring emerging phenomena  across the research fronts of spectroscopy, quantum optics, plasmonics, and polaritonics. 
    \end{abstract}

	\maketitle

    \section{Introduction}
    
    The intersection of quantum chemistry and nanophotonics has opened new frontiers in the study of light-matter interactions, ranging from plasmon-enhanced spectroscopy \cite{Odom2011,Aroca2013,Wang2020NatRevPhys} to the burgeoning field of polariton chemistry \cite{Ribeiro2018,Flick2018,Herrera2019,Li2022Review,Fregoni2022,Simpkins2023,Mandal2023ChemRev,Ruggenthaler2023,Xiang2024}. Numerically modeling these systems presents fundamental challenges: one must accurately capture the quantum dynamics of microscopic molecules while simultaneously describing the behavior of the electromagnetic (EM) field as it interacts with complex, micro-scale geometries. While powerful simulation tools exist for each domain --- notably the finite-difference time-domain (FDTD) method \cite{Taflove2005,Oskooi2010} for classical electrodynamics and numerous codes in quantum chemistry and computational physics for propagating molecular dynamics across different time and length scales \cite{Johansson2012,HjorthLarsen2017,Thompson2022,Smith2020,Litman2024} --- these approaches have largely been developed in separate ecosystems.
    
    Efforts have been made to propagate coupled EM-molecular dynamics self-consistently \cite{Sukharev2017,Luk2017,Li2018Spontaneous,Hoffmann2018,Li2020Water,Sukharev2023a,Xu2023Polariton,Zhou2024,Bonafe2025}. In such schemes, microscopic molecules radiate into the EM field, while local EM environments simultaneously modulate molecular dynamics via light-matter coupling. Depending on the level of theory used to describe the microscopic molecules, a hierarchy of self-consistent schemes has been established, such as coupling Maxwell's equations to optical Bloch equations \cite{Castin1995}, time-dependent density functional theory (TDDFT) \cite{Lopata2009-1,Lopata2009-2,Chen2010,Bonafe2025,Sidler2025}, or classical and nonadiabatic molecular dynamics \cite{Yamada2020,Li2024CavMD,Sokolovskii2022tmp,Nguyen2025}. Many previous implementations, however, possess limited modeling capacity for either the EM field or the microscopic molecules, thereby only partially exploiting state-of-the-art computational developments within the EM and molecular simulation ecosystems. Additionally, these implementations typically require deep, intrusive modifications to the source code of the EM or molecular solvers, resulting in monolithic codes that are difficult to maintain and extend.

    Here, we report the development of \MaxwellLink, a flexible and extensible framework for coupling classical EM solvers to external molecular drivers. \MaxwellLink is designed as a universal EM-molecular engine, allowing theoretical chemists and computational physicists to readily connect their preferred molecular simulation codes  \cite{Johansson2012,HjorthLarsen2017,Thompson2022,Smith2020} to a variety of EM solvers with minimal glue code. In contrast to routine EM simulations \cite{Taflove2005,Oskooi2010}, \MaxwellLink upgrades the description of materials from conventional dielectric functions to a variety of advanced molecular engines.

    A unique feature of \MaxwellLink is its ability to connect an EM solver, such as the finite-difference time-domain (FDTD) approach, to many \textit{heterogeneous} molecular drivers concurrently (i.e., molecules described under different levels of theory). The EM solver and molecular drivers communicate through a uniform TCP/UNIX socket interface  that is robust against the termination and reconnection of the molecular drivers. Leveraging this socket interface, \MaxwellLink enables the simultaneous parallel acceleration of the EM engine and molecular drivers on separate computational nodes or even distinct high-performance computing (HPC) systems, an architecture that is particularly appealing for large-scale light-matter simulations. For demonstrative calculations, the EM solver and molecular drivers can also connect locally within the same processor in the absence of the socket communication.
    
    In this initial work, \MaxwellLink incorporates three classical EM solvers (Table \ref{table:em_solvers}) and six molecular drivers (Table \ref{table:molecular_drivers}). The EM solvers range from (i) MEEP, an industry-standard open-source FDTD engine \cite{Oskooi2010}, to (ii) a classical single-mode cavity, and (iii) arbitrary laser fields for pumping molecules. The molecular drivers include model systems, such as (i) a lightweight two-level system (TLS) driver and (ii) a QuTiP  interface \cite{Johansson2012} for custom model Hamiltonians with optional Lindblad dissipation; electronic ground-state molecular mechanics, such as (iii) first-principles Born--Oppenheimer molecular dynamics (MD) via the ASE interface \cite{HjorthLarsen2017}, and (iv) classical force-field molecular dynamics utilizing the LAMMPS package \cite{Thompson2022} with an embedded socket interface; and first-principles nonadiabatic quantum dynamics models, such as in-house (v) real-time time-dependent density functional theory (RT-TDDFT) \cite{Bruner2016,Dar2024} and (vi) real-time Ehrenfest (RT-Ehrenfest) dynamics \cite{Li2005Eh,Goings2018} using electron integrals from the Psi4 quantum chemistry package \cite{Smith2020}. As an open-source Python project, \MaxwellLink also includes a detailed documentation website \cite{MaxwellLinkDocument} to facilitate the adoption of this code by newcomers and students.

    By integrating these EM solvers and molecular drivers, \MaxwellLink establishes a uniform simulation platform for light-matter dynamics. Within the same Python interface, users gain the flexibility to switch levels of theory for both the EM solvers and molecular drivers, proceeding from simplified models to realistic large-scale calculations. This flexibility may potentially catalyze the exploration of different research fronts in light-matter interactions, which typically adopt a particular level of theory for describing either the light or matter component. Because the \MaxwellLink framework greatly simplifies the procedure and flattens the learning curve of self-consistent light-matter simulations, future method development for more accurate and efficient light-matter simulations also becomes more approachable.

    To demonstrate the capabilities of \MaxwellLink, we provide four examples ranging from model systems to realistic calculations: (i) superradiance \cite{Dicke1954,Gross1982} of a large collection of TLSs in two-dimensional (2D) vacuum; (ii) radiative energy transfer from a TLS donor to an HCN molecule acceptor described by different levels of theory in three-dimensional (3D) vacuum; (iii) vibrational strong coupling of liquid water confined in a single-mode cavity \cite{Li2020Water} versus a one-dimensional (1D) Bragg resonator; and (iv) vibrational heating of a few hundred gas-phase HCN molecules near a 3D plasmonic metamaterial. These four examples not only offer a glimpse into the scientific problems \MaxwellLink can address, but also highlight unique features of the code, such as the concurrent connection with a large collection of molecular drivers, the flexibility of switching levels of theory for both the EM field and molecules, and the parallel acceleration of the EM solver and molecular drivers.
    
    This manuscript is organized as follows. Sec. \ref{sec:theory} reviews various numerical frameworks for simulating both light and matter and provides a unified description of these frameworks within \MaxwellLink. Sec. \ref{sec:implementation_details} introduces the implementation details and code structure of the package. Sec. \ref{sec:results} presents selected demonstrative calculations. Finally, Sec. \ref{sec:conclusion} concludes this work and outlines the future development of \MaxwellLink.

    \section{Theory}\label{sec:theory}

    In this section, we outline different levels of theory for  self-consistent light-matter dynamics implemented in \MaxwellLink.

    \subsection{Schemes for propagating classical EM dynamics}

    Under two fundamental assumptions --- (i) a classical description of the EM field and (ii) restriction to electric-field coupling with molecules --- the propagation of the EM field  in the time domain falls into three primary categories: simulating Maxwell's equations (i) on a real-space grid, (ii) within a normal-mode basis, or (iii) by directly applying analytical solutions of the Maxwell's equations under simplified conditions. Below, we  demonstrate that all the three EM propagation schemes can be uniformly integrated into the \MaxwellLink framework. 
    
    \subsubsection{Classical Maxwell's equations on a spatial grid}
        
    We begin with the classical Maxwell's equations: \cite{Griffiths1999}
    \begin{subequations}\label{eq:Maxwell}
        \begin{align}
\partial_t \vD (\vR,t) & =\vnabla \times \mathbf{H}(\mathbf{r},t) - \vJ_{\text{mol}}(\vR, t) , \\
    \partial_t \vB(\mathbf{r},t) &= - \vnabla \times \vE (\vR,t) .
\end{align}
    \end{subequations}
    Here, the displacement field $\vD$ is related to the electric field $\vE$ via 
    \begin{equation}\label{eq:Dromega}
        \vD(\vR, \omega) = \epsilon_0 \epsilon_{\rm r}^{\ast}(\vR, \omega) \vE(\vR, \omega),
    \end{equation}
    where $\epsilon_0$ denotes the vacuum permittivity, and the dimensionless quantity $\epsilon_{\rm r}^{\ast}(\vR, \omega)$ represents the relative  permittivity, or dielectric function, of the classical media. In FDTD, dielectric functions are frequently expressed using a linear combination of Lorentz and Drude oscillators with parameters fitted from experiments.  As Eq. \eqref{eq:Dromega} is defined in the frequency domain, updating $\vE(\vR, t)$ requires a convolution of $\vD(\vR, t')$ during prior time steps $t' < t$. Similarly, the magnetic $\vB$ field is related to the magnetic $\vH$ field  via $\vB = \mu \vH$. For non-magnetic materials, $\mu = \mu_0$, the vacuum magnetic permeability. The propagation of $\vD$, $\vH$, $\vE$, and $\vB$ fields, and the representation of the matter by classical dielectric functions, can be efficiently solved using standard FDTD algorithms, which simulate EM fields in staggered temporal-spatial grid, also known as the Yee's cell \cite{Yee1966}.

    Using classical dielectric functions to represent the matter component is generally sufficient for describing linear optical spectroscopy. Beyond this regime, it is often desirable to directly include realistic microscopic molecules in Maxwell's equations. In our framework, the microscopic molecular response to the EM field is encoded in the classical molecular current density $\vJ_{\text{mol}}(\vR, t)$,  defined as \cite{Sukharev2017,Li2018Spontaneous}
    \begin{equation}\label{eq:J_mol}
        \vJ_{\text{mol}}(\vR, t) = \sum_{m=1}^{N_{\text{mol}}} \partial_t \vP_{\text{mol}}^m(\vR, t) ,
    \end{equation}    
    where $\vP_{\text{mol}}^m(\vR, t)$ represents the classical polarization density of molecule $m$.

    Having introduced the classical Maxwell's equations, we proceed to evaluate the classical polarization density of each microscopic molecular site, $\vP_{\text{mol}}^m(\vR, t)$. We express this quantity as follows:
    \begin{equation}\label{eq:Polarization_mol}
        \vP_{\text{mol}}^m(\vR, t) =  \sum_{i=x,y,z} \gamma \mu_m^i(t) \vkappa_m^{i}(\vR) ,
    \end{equation}
    where $\mu_m^i$ represents the classical dipole moment of the $m$-th molecular site along direction $i=x,y,z$, evaluated under the point-dipole approximation, and $\gamma$ is the rescaling factor of light-matter coupling ($\gamma=1$ by default). The spatial kernel function $\vkappa_m^{i}(\vR)$ is oriented along direction $i=x,y,z$, and is normalized over the volume of molecular distribution $\Omega_m$: $ \left |\int_{\Omega_m} d\vR \ \vkappa_m^{i}(\vR) \right | = 1$.
    Obviously, the kernel function $\vkappa_m^{i}(\vR)$ characterizes the spatial distribution of molecular polarization density. While the detailed form of the kernel function depends on the problem of interest, it is frequently set to a Gaussian function with a tunable width in our simulations.
    
    Although the molecular size and polarization distribution are typically orders of magnitude smaller than that of the EM wavelength, we adopt a strategy in which the spatial width of the kernel function $\vkappa_m^{i}(\vR)$ is set approximately one-tenth of the relevant EM wavelength, being the same order of magnitude as the spatial resolution of FDTD simulations. The goal of this strategy is twofold. On the one hand, this strategy still ensures the point dipole approximation for microscopic molecules with respect to the EM wavelength, while avoiding the use of a very fine spatial grid in FDTD simulations; on the other hand, as shown immediately below, it effectively smooths the self-emitted EM field compared to the  treatment of assigning the molecular polarization on a single grid point, thus avoiding dealing with numerical singularities when evaluating the light-matter coupling.

    Given the molecular polarization density in Eq. \eqref{eq:Polarization_mol},  standard EM theory defines the classical light-matter coupling for molecule $m$ as
    \begin{equation}\label{eq:Vm_standard}
        V_m = -\int_{\Omega_m} d\vR\ \vE(\vR, t)\cdot \vP_{\text{mol}}^{m} (\vR, t) .
    \end{equation}
    Here, the electric field encompasses both the longitudinal ($\vE_{\parallel}$) and transverse ($\vE_{\perp}$)  components ($\vE = \vE_{\parallel} + \vE_{\perp}$) \cite{Cohen-Tannoudji1997}, i.e., both the photonic (from $\vE_{\perp}$) and electrostatic interactions (from $\vE_{\parallel}$, such as dipole-dipole coupling) between the environment and the local molecular site are treated at the level of classical electric fields. Conversely, short-range interactions internal to a specific molecular site remain accurately described by the specific level of molecular theory employed.
    
    Then, by substituting the definition of $\vP_{\text{mol}}^{m} (\vR, t) $ in Eq. \eqref{eq:Polarization_mol} into Eq. \eqref{eq:Vm_standard}, we obtain a simpler form of the light-matter coupling:
    \begin{subequations}\label{eq:E_regu}
        \begin{equation}\label{eq:E_regu_vm}
        V_m  = -\sum_{i=x,y,z}  \mu_m^i(t) \widetilde{E}_m^i(t) .       
    \end{equation}
    In Eq. \eqref{eq:E_regu_vm}, the regularized electric field amplitude $\widetilde{E}_m^i(t)$ is defined as
    \begin{equation}\label{eq:E_regu_Emi}
        \widetilde{E}_m^i(t) \equiv \int_{\Omega_m} d\vR\   \vE(\vR, t) \cdot \gamma \vkappa^{i}_m(\vR) .
    \end{equation}
    In other words, molecular dipole $m$ now effectively experiences a regularized electric field vector
    \begin{equation}\label{eq:E_regu_Em}
        \widetilde{\vE}_m(t) = [\widetilde{E}_m^x(t), \widetilde{E}_m^y(t), \widetilde{E}_m^z(t)] ,
    \end{equation}
    with $\widetilde{E}_m^{x,y,z}(t)$ defined in Eq. \eqref{eq:E_regu_Emi}.
    
    Since $\widetilde{\vE}_m(t)$ in Eq. \eqref{eq:E_regu_Em} is computed by weighting the raw electric field $\vE(\vR, t)$ with the spatial kernel function $\vkappa^{i}_m(\vR)$, $\widetilde{\vE}_m(t)$ does not inherit the singular spatial fluctuations in $\vE(\vR, t)$, thereby avoiding numerical instabilities in coupled EM-molecular simulations.
    \end{subequations}
    
    From a practical perspective, introducing the regularized electric field $\widetilde{\vE}_m(t)$ proves advantageous  for EM-molecular communication. As the  regularized electric field vector $\widetilde{\vE}_m(t)$ can be efficiently computed within the FDTD engine, \MaxwellLink needs only send the  three-component vector  $\widetilde{\vE}_m(t)$ to each molecular driver. In return, each molecular driver then transmits a three-component vector $\dot{\vmu}_m = [\dot{\mu}_m^x, \dot{\mu}_m^y, \dot{\mu}_m^z]$ for constructing the molecular current density in FDTD [cf. Eqs. \eqref{eq:J_mol} and \eqref{eq:Polarization_mol}]. 
    Given this minimal inter-code communication overhead, \MaxwellLink is capable of efficiently simulating a large number of molecular sites interacting with the FDTD engine.

    \subsubsection{Classical photonic dynamics in normal modes}

    An alternative strategy for simulating the EM field involves propagating classical photonic dynamics in a normal-mode basis. This method avoids constructing the EM field on a real-space grid; however, it is typically practical only for simplified boundary conditions where the photonic mode functions can be expressed either analytically or semi-analytically.

    For example, a recent mesoscale cavity molecular dynamics (CavMD) \cite{Li2024CavMD} scheme utilizes the following field Hamiltonian:
    \begin{equation}\label{eq:H_F_normalmode}
        \hat{H}_{\rm F} = \sum_{k\lambda} \frac{1}{2} p_{k\lambda}^2 + \frac{1}{2}\omega_{k}^2 \left(
        q_{k\lambda} + \sum_{m} \frac{\hat{\vmu}^{(m)}\cdot \mathbf{f}_{k\lambda}(\vR_m)}{\sqrt{\epsilon_0}\omega_{k}}
        \right)^2 .
    \end{equation}
    Here, $p_{k\lambda}$, $q_{k\lambda}$, and $\omega_{k}$ denote the momentum, position, and frequency of a photonic normal mode defined by wave vector $k = |\vk|$ and polarization vector $\vxi_{\lambda}$ (satisfying $\vxi_{\lambda} \cdot \vk = 0$). $\hat{\vmu}^{(m)}$ is the dipole operator of molecular site $m$, and $\mathbf{f}_{k\lambda}(\vR_m)$ represents the value of the photonic mode function at the position of this molecular site ($\vR_m$).

    With the classical propagation of the field Hamiltonian in Eq. \eqref{eq:H_F_normalmode}, each photonic normal mode is governed by the following  Newtonian equation of motion:
    \begin{subequations}\label{eq:eom_cavmd}
    \begin{equation}\label{eq:eom_cavmd_a}
        \ddot{q}_{k\lambda} = -\omega_{k}^2 q_{k\lambda} - \varepsilon_{k\lambda} d_{k\lambda} .
    \end{equation}
    In Eq. \eqref{eq:eom_cavmd_a}, the effective light-matter coupling $\varepsilon_{k\lambda}$ is defined as
    \begin{equation}\label{eq:varepsilon_cavmd}
        \varepsilon_{k\lambda} \equiv \frac{\omega_{k}}{\sqrt{\mathcal{V}\epsilon_0}} ,
    \end{equation}
    where $\mathcal{V}$ denotes the effective volume of the photonic environment. The total molecular dipole moment $d_{k\lambda}$ experienced by photonic mode $k\lambda$ is then evaluated by
    \begin{equation}\label{eq:dklambda_def}
        d_{k\lambda} \equiv \sum_{m} \sqrt{\mathcal{V}} \avg{\hat{\vmu}^{(m)}} \cdot \mathbf{f}_{k\lambda}(\vR_m) ,
    \end{equation}
    where $\avg{\cdots}$ denotes the molecular mean-field average.
    \end{subequations}

    By rewriting the light-matter coupling in Hamiltonian \eqref{eq:H_F_normalmode} as $-\sum_m \widetilde{\vE}_m\cdot \hat{\vmu}^{(m)}$ and differentiating Hamiltonian \eqref{eq:H_F_normalmode} with respect to $\hat{\vmu}^{(m)}$, we obtain the effective classical electric field vector $\widetilde{\vE}_m(t)$ experienced by molecular site $m$:
    \begin{equation}\label{eq:E_normalmode}
        \widetilde{\vE}_m(t) = - \sum_{k\lambda} 
       \left[ \varepsilon_{k\lambda} q_{k\lambda}(t) + \frac{\varepsilon_{k\lambda}^2}{\omega_k^2} d_{k\lambda}(t) \right] \sqrt{\mathcal{V}} \mathbf{f}_{k\lambda}(\vR_m).
    \end{equation}
    Here, $\varepsilon_{k\lambda}$ and $d_{k\lambda}$ have been defined in Eq. \eqref{eq:varepsilon_cavmd} and Eq. \eqref{eq:dklambda_def}, respectively. On the right-hand side of Eq. \eqref{eq:E_normalmode},
    the term containing the $\varepsilon_{k\lambda}^2$ factor  arises from the dipole self-energy term in Hamiltonian \eqref{eq:H_F_normalmode}. 

    In the single-mode limit, assuming a cavity photon mode polarized along both the $x$- and $y$-directions, the photonic mode function can be set as $\mathbf{f}_{k\lambda} = \sqrt{1/\mathcal{V}}\ve_\lambda$, where $\lambda =x,y$ and $\ve_\lambda$ denotes the corresponding unit vector. The current version of \MaxwellLink implements photonic normal-mode dynamics  under the single-mode limit only, although extending the implementation to normal-mode dynamics beyond this limit is straightforward.

    It is evident that in the normal-mode basis, each molecular site $m$ requires only the local 3D electric field vector $\widetilde{\vE}_m$, similar to the FDTD case. Each photonic normal mode requires $d_{k\lambda}$ [Eq. \eqref{eq:dklambda_def}], the total dipole moment of all molecular sites weighted by the photonic mode functions; this quantity can be constructed on the fly once the EM engine gathers the dipole moment vectors of all molecular sites. Hence, in both the FDTD and normal-mode cases, the communication load between the EM engines and each molecular driver becomes exactly the same, thereby enabling the uniform implementation of these two schemes within the \MaxwellLink framework.

    \subsubsection{Using analytical solutions of Maxwell's equations}

    For EM fields interacting with materials, Maxwell's equations in the time domain can be formally solved, leading to Jefimenko's equations \cite{Griffiths1999}. For quantum emitters in a dielectric medium, the radiative interaction between the EM field and local quantum emitters can be evaluated analytically using the dyadic Green's function approach \cite{Novotny2006,Ding2017}. For a classical emitter in vacuum, it is also well established that the radiative self-interaction can be represented analytically as the Abraham--Lorentz force \cite{Griffiths1999}.

    While these analytical results account for the back-action from the molecules to the EM field, a simpler treatment ignores this back-action. This approximation, sometimes known as the classical path approximation \cite{Smith1969}, is widely applied in the study of light-matter interactions. Under this approximation, for example, in 1D vacuum a simple wave form of the electric field can be used to excite molecules:
    \begin{equation}\label{eq:E_analytical}
        \widetilde{\vE}(z, t) = \widetilde{\vE}(z -ct),
    \end{equation}
    assuming propagation along the $z$-axis.
    Naturally, analytical solutions of Maxwell's equations can also be integrated into the \MaxwellLink framework, where a local three-dimensional E-field vector $\widetilde{\vE}(z, t)$ is required to propagate a molecular site.    

    Table \ref{table:em_solvers} summarizes the three EM solvers available in the initial version of \MaxwellLink, each representing a numerical scheme introduced above for propagating the EM dynamics. Note that Coulomb-gauge light-matter dynamics, i.e., charged particles interacting with the EM vector potential, are not supported in this initial version of \MaxwellLink.

    \begin{table*}
    \caption{Available EM solvers in \MaxwellLink.
    }
    \label{table:em_solvers}
    \begin{tabular}{lcr}
    \hline
    \hline
    EM solvers \ \ & Equations of motion \ \  & Description \\
    \hline
    \texttt{MeepSimulation} & Eq. \eqref{eq:Maxwell}  & Full-feature FDTD solver by interfacing MEEP  \\
    \texttt{SingleModeSimulation} & Eqs. \eqref{eq:H_F_normalmode}-\eqref{eq:E_normalmode} & A single cavity mode for simplified EM dynamics \\
    \texttt{LaserDrivenSimulation} & Eq. \eqref{eq:E_analytical}  & Custom electric fields, no back-action from molecules \\
    \hline
    \hline
    \end{tabular}
    \end{table*}

    \begin{table*}
    \caption{Available molecular drivers in \MaxwellLink. \footnote{All molecular drivers support both the embedded (non-socket) and socket modes except \texttt{lammps}.}
    }
    \label{table:molecular_drivers}
    \begin{tabular}{lcr}
    \hline
    \hline
    Molecular drivers \ \ & Equations of motion \ \  & Description \\
    \hline
    \texttt{tls} & Eq. \eqref{eq:von-Neumann-tls}  & Lightweight driver for a TLS \\
    \texttt{qutip} & Eq. \eqref{eq:von-Neumann-tls} & Interfacing QuTiP for custom model Hamiltonians \\
    \texttt{ase} & Eq. \eqref{eq:F_MM} & Interfacing ASE for first-principles Born--Oppenheimer MD \\
    \texttt{lammps} & Eq. \eqref{eq:F_MM} & Modified LAMMPS code (socket mode only) \\
    \texttt{rttddft} & Eq. \eqref{eq:von-Neumann-rtTDDFT} & In-house code for RT-TDDFT using Psi4 integrals \\
    \texttt{rtehrenfest} & Eqs. \eqref{eq:von-Neumann-rtTDDFT} \& \eqref{eq:F_ehrenfest}  & In-house code for RT-Ehrenfest dynamics using Psi4 integrals \\
    \hline
    \hline
    \end{tabular}
    \end{table*}

    \subsection{Molecular dynamics under EM fields}

    The above introduces three schemes for propagating EM fields that can be uniformly integrated within the \MaxwellLink framework. We now outline three different classes of molecular equations of motion available in the initial version of \MaxwellLink: (i) model Hamiltonians; (ii) classical or Born--Oppenheimer molecular mechanics; and (iii) first-principles electronic nonadiabatic molecular dynamics. For simplicity, we focus on the description of a single molecular site below, thereby omitting the index $m$ used to distinguish between different molecular sites.

    \subsubsection{Model quantum systems}
    \MaxwellLink supports two types of model quantum systems: (i) a simple two-level system (TLS), and (ii) arbitrary open quantum model Hamiltonians via an interface with the QuTiP package \cite{Johansson2012}. In both cases, the propagation of model Hamiltonians follows the standard von Neumann equation:
    \begin{equation}\label{eq:von-Neumann-tls}
        \frac{d}{dt}\hat{\rho}(t) = -\frac{i}{\hbar} \left [\hH - \widetilde{\vE}(t)\cdot \hat{\vmu}, \  \hat{\rho}(t) \right ] - \mathcal{L}[\hat{\rho}] .
    \end{equation}
    In the case of the TLS driver, the field-free Hamiltonian is given by $\hH = \bigl( \begin{smallmatrix}0 & 0\\ 0 & \hbar\omega_0\end{smallmatrix}\bigr)$, where $\omega_0$ is the transition frequency. The transition dipole moment operator $\hat{\vmu}$ is defined as $\hat{\vmu} =  \mu_{01} \ve_i\hat{\sigma}_x $, where $\mu_{01}$ denotes the magnitude of the transition dipole moment, the unit vector $\ve_i$ ($i=x,y,z$) defines the orientation of the dipole vector, and $\hat{\sigma}_x = \bigl( \begin{smallmatrix}0 & 1\\ 1 & 0\end{smallmatrix}\bigr)$ represents the Pauli-$x$ matrix. Note that Lindblad dissipation $\mathcal{L}[\hat{\rho}]$ is not supported in the basic TLS driver. When utilizing the QuTiP interface, users must supply a custom $\hH$, $\hat{\vmu}$, and optionally $\mathcal{L}[\hat{\rho}]$ to propagate arbitrary open model Hamiltonians. 

    At each EM time step, following the propagation of Eq. \eqref{eq:von-Neumann-tls}, the time derivative of the dipole moment expectation value is calculated:
    \begin{equation}\label{eq:dipole_derivative}
        \frac{d}{dt} \avg{\hat{\vmu}} = \tr{\frac{d}{dt}\hat{\rho} \hat{\vmu}} .
    \end{equation}
    This quantity is required to evolve Maxwell's equations [cf. Eqs. \eqref{eq:J_mol} and \eqref{eq:Polarization_mol}] using FDTD.
    For the TLS, the analytical form $\frac{d}{dt} \avg{\hat{\vmu}} = -2\omega_0 \mu_{01}\text{Im}(\rho_{01})\ve_i$ is derived by substituting Eq. \eqref{eq:von-Neumann-tls} into the trace in Eq. \eqref{eq:dipole_derivative},
    where $\text{Im}(\rho_{01})$ denotes the imaginary component of the off-diagonal coherence in the density operator. Additionally, $\avg{\hat{\vmu}}$ is computed for propagating photonic dynamics under the normal-mode basis.

    \subsubsection{Classical Born--Oppenheimer MD}

    The second class of molecular drivers for \MaxwellLink enables the coupling of Maxwell's equations to classical Born--Oppenheimer MD. Generally, when molecules evolve on the electronic ground-state potential energy surface, the classical nuclear Hamiltonian is written as:
    \begin{equation}\label{eq:H_BO}
        H = H_0(\{\mathbf{P}_n\}, \{\mathbf{R}_n\}) - \vmu \cdot \widetilde{\vE}(t) .
    \end{equation}
    In Eq. \eqref{eq:H_BO}, $H_0(\{\mathbf{P}_n\}, \{\mathbf{R}_n\})$ represents the standard field-free (kinetic + potential) nuclear Hamiltonian, which depends on the momenta ($\{\mathbf{P}_n\}$) and coordinates ($\{\mathbf{R}_n\}$) of all atoms; $\vmu$ denotes the classical molecular dipole vector, calculated by
    \begin{equation}\label{eq:dipole_partial_charge}
        \vmu = \sum_{n} Q_n \mathbf{R}_n,
    \end{equation}
    where $Q_n$ is the partial charge of the $n$-th atom.

    In practice, the interaction with the external field is implemented by modifying the molecular forces at each time step:
    \begin{equation}\label{eq:F_MM}
        \mathbf{F}_n^{\text{new}}(t) = \mathbf{F}_n^{\text{old}}(t) + Q_n \widetilde{\vE}(t) .
    \end{equation}
    These updated nuclear forces are then employed to propagate the classical MD. 
    
    To communicate with the FDTD engine, the molecular driver transmits the time derivative of the dipole vector at each time step:
    \begin{equation}\label{eq:dvmu_BO}
        \dot{\vmu} = \sum_{n} Q_n \mathbf{V}_n ,
    \end{equation}
    where $\mathbf{V}_n$ denotes the velocity of the $n$-th atom. For communication with the normal-mode photonic engine, apart from sending the dipole time derivative in Eq. \eqref{eq:dvmu_BO}, the molecular drivers also calculate the dipole vector ($\vmu$) of each molecular site.

    Two molecular drivers are available in this category: (i) first-principles Born--Oppenheimer MD supported by the ASE Python interface \cite{HjorthLarsen2017}; and (ii) classical empirical or machine-learning force-field MD achieved by directly connecting to the open-source LAMMPS code \cite{Thompson2022}. 
    
    \subsubsection{RT-TDDFT and RT-Ehrenfest dynamics}

    Beyond model systems and Born--Oppenheimer MD, the electronic dynamics of realistic molecules can be propagated via real-time electronic structure theory, i.e., by solving the time-dependent Schr\"odinger equation for many-electron molecular systems. Within the class of real-time electronic structure methods, the RT-TDDFT approach is widely utilized due to its computational efficiency \cite{Bruner2016,Dar2024}. 

    Analogous to the model systems, RT-TDDFT propagates the von Neumann equation for the electronic density matrix $\mathbf{P}^{\rm e}_{\rm o}$:
    \begin{equation}\label{eq:von-Neumann-rtTDDFT}
       \frac{d}{dt} \mathbf{P}^{\rm e}_{\rm o}(t) = -\frac{i}{\hbar} \left [\mathbf{F}^{\rm e}_{\rm o}(t) - \widetilde{\vE}(t)\cdot \vec{\vmu}^{\rm e}_{\rm o}, \  \mathbf{P}^{\rm e}_{\rm o}(t) \right ] .
    \end{equation}
    Here, $\mathbf{F}^{\rm e}_{\rm o}$ represents the closed-shell electronic Kohn--Sham matrix, and $\vec{\vmu}^{\rm e}_{\rm o} \equiv (\vmu^{\text{e}x}_{\rm o}, \vmu^{\text{e}y}_{\rm o}, \vmu^{\text{e}z}_{\rm o} )$ denotes the vector of electronic dipole matrices along the Cartesian coordinates. While the von Neumann equation is propagated in the orthogonal atomic orbital basis (denoted by the subscript $_{\rm o}$), the electronic Kohn--Sham matrix must be constructed in the non-orthogonal atomic orbital basis (denoted by the absence of the subscript $_{\rm o}$):
    \begin{equation}\label{eq:Fock_RTTDDFT}
    \begin{aligned}
        \mathbf{F}^{\rm e}(t) & = \mathbf{H}_{\rm{core}}^{\rm e} + \mathbf{J}^{\rm ee}[\mathbf{P}^{\rm e}(t)] + \zeta \mathbf{K}^{\rm ee}[\mathbf{P}^{\rm e}(t)] \\
        & + (1-\zeta) \mathbf{V}_{\rm xc}^{\rm e} [\mathbf{P}^{\rm e}(t)].
    \end{aligned}
    \end{equation}
    Here, $\mathbf{H}_{\rm{core}}^{\rm e}$ denotes the core Hamiltonian, which includes the electronic kinetic energy and the electrostatic interaction with nuclei (treated as fixed point charges); $\mathbf{J}^{\rm ee}$, $\mathbf{K}^{\rm ee}$, and $\mathbf{V}_{\rm xc}^{\rm e}$ represent the electronic Coulomb interaction, Hartree--Fock exchange interaction, and the exchange-correlation potential under the adiabatic approximation, respectively. The parameter $0 \leq \zeta \leq 1$ interpolates the theory between the Hartree--Fock limit ($\zeta = 1$) and the pure density functional theory (DFT) limit ($\zeta = 0$). Hybrid DFT functionals employ a value of $0 < \zeta < 1$. Because the $\mathbf{J}^{\rm ee}$, $\mathbf{K}^{\rm ee}$, and $\mathbf{V}_{\rm xc}^{\rm e}$ terms in Eq. \eqref{eq:Fock_RTTDDFT} depend on the electronic density matrix, $\mathbf{F}^{\rm e}(t)$ must be reconstructed at each time step, a procedure distinct from that adopted in model system calculations.

    Since the electronic Kohn--Sham matrix is constructed and propagated in different bases, a basis transformation between the non-orthogonal and orthogonal atomic orbital bases is required at each time step:
    \begin{subequations}\label{eq:basis_transfer_rttddft}
        \begin{align}
            \mathbf{P}^{\rm e}_{\rm o} &= [\mathbf{S}^{\rm e}]^{1/2} \mathbf{P}^{\rm e} [\mathbf{S}^{\rm e}]^{1/2} , \\
            \mathbf{F}^{\rm e}_{\rm o} &= [\mathbf{S}^{\rm e}]^{-1/2} \mathbf{F}^{\rm e} [\mathbf{S}^{\rm e}]^{-1/2} ,
        \end{align}
    \end{subequations}
    where $\mathbf{S}^{\rm e}$ denotes the electronic overlap matrix, which is generally not an identity matrix when localized Gaussian atomic-orbital basis functions are used. The basis transformation for the dipole matrix vector $\vec{\vmu}_{\rm o}^{\rm e}$ follows the same procedure as that for $\mathbf{F}^{\rm e}_{\rm o}$ [Eq. \eqref{eq:basis_transfer_rttddft}].

    In conjunction with the RT-TDDFT electronic dynamics, classical nuclear dynamics can be propagated on a mean-field electronic surface, leading to the RT-Ehrenfest scheme \cite{Li2005Eh,Goings2018,Zhao2020JCP,Li2023eBO}. The associated mean-field nuclear force is defined by
    \begin{equation}\label{eq:F_ehrenfest}
        \mathbf{F}_n = -\vnabla_n U[\mathbf{P^{\rm{e}}}(t), \{ \mathbf{R}_n\} ] + Z_n \widetilde{\mathbf{E}}_n .
    \end{equation}
    where $U[\mathbf{P^{\rm{e}}}(t), \{ \mathbf{R}_n\} ]$ represents the molecular energy associated with the Fock matrix $\mathbf{F}^{\rm e}(t) - \widetilde{\vE}(t)\cdot \vec{\vmu}^{\rm e}$, and $Z_n$ denotes the atomic number of nucleus $n$. Here, both the electronic and nuclear responses to the electric field are accounted for in the force calculations.

    To propagate Maxwell's equations using FDTD [cf. Eqs. \eqref{eq:J_mol} and \eqref{eq:Polarization_mol}], we also calculate the time derivative of the dipole moment expectation value at each time step:
    \begin{equation}\label{eq:dipole_derivative_rtTDDFT}
        \frac{d}{dt} \avg{\hat{\vmu}} = 2\tr{\frac{d}{dt}\mathbf{P}^{\rm e}(t) \vec{\vmu}^{\text{e}}  } + \sum_n Z_n \mathbf{V}_n,
    \end{equation}
    where $\frac{d}{dt}\mathbf{P}^{\rm e}(t)$ is calculated using Eq. \eqref{eq:von-Neumann-rtTDDFT}, and $\mathbf{V}_n$ denotes the velocity of each nucleus. The prefactor 2 in Eq. \eqref{eq:dipole_derivative_rtTDDFT} accounts for both the $\alpha$ and $\beta$ electrons in restricted Kohn--Sham calculations. The negative electronic charge is incorporated into the definition of $\vec{\vmu}^{\text{e}}$. When communicating with normal-mode photonic dynamics, $\avg{\hat{\vmu}}$ for each molecular site is also computed.
 
    In both the RT-TDDFT and RT-Ehrenfest Python drivers, electronic matrices are constructed using electron integrals from the Psi4 quantum chemistry package \cite{Smith2020}. This open-source reference implementation serves as a benchmark baseline for future integration between \MaxwellLink and more efficient electronic structure packages, which are primarily closed-source at present.

    Table \ref{table:molecular_drivers} summarizes the six molecular drivers available in \MaxwellLink.

    \section{Implementation details}\label{sec:implementation_details}

    \begin{figure*}
		\centering
		\includegraphics[width=0.75\linewidth]{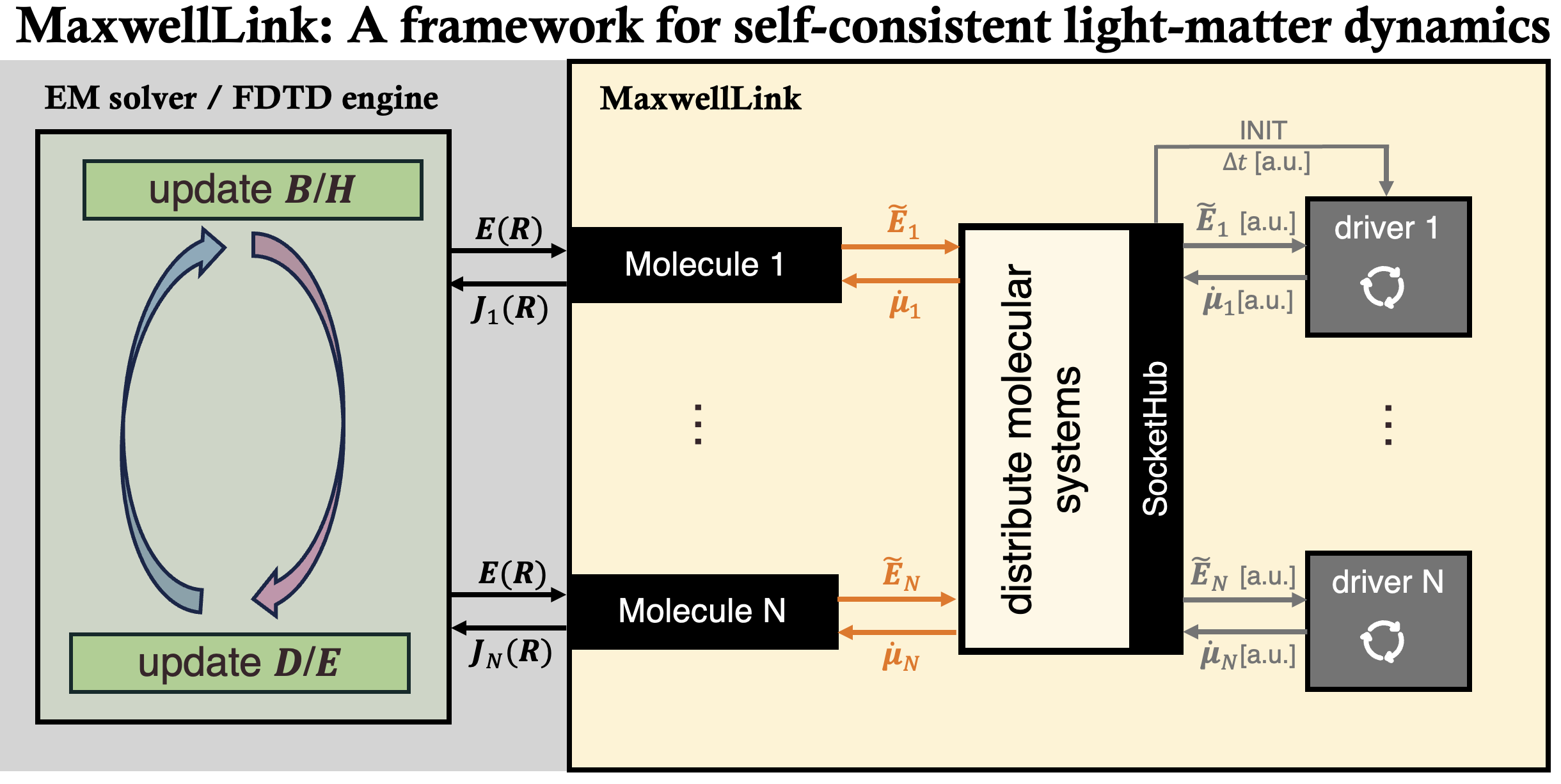}
		\caption{ 
        \textbf{Modular design in the \MaxwellLink package for self-consistent EM-molecular simulations.} The EM solver interfaces with various abstract \texttt{Molecule} instances, which store only EM-relavent information such as the molecular location and size within a real-space EM grid. These abstract \texttt{Molecule} instances communicate with external molecular drivers via a \texttt{SocketHub} instance using the TCP/UNIX socket protocol. The corresponding Python input for launching \MaxwellLink simulations is provided in Code Listing \ref{code:sample}. Python molecular drivers can also attach directly to the abstract \texttt{Molecule} instances without employing the socket interface.
		}
		\label{fig:workflow}
    \end{figure*}

    After introducing the fundamental theory, we now provide an overview of the \MaxwellLink framework. To provide a universal light-matter engine, \MaxwellLink adopts a modular design principle. As shown in Fig. \ref{fig:workflow}, the EM solver module interacts with abstract \texttt{Molecule} instances, which are agnostic to the level of theory used to describe the underlying molecules. Instead, these instances store only EM-relevant information, such as the molecular coordinates in real EM space and the spatial kernel function required for calculating the regularized electric field vector (in FDTD solvers). All \texttt{Molecule} instances communicate concurrently with a \texttt{SocketHub} instance, which manages the socket interface between the abstract \texttt{Molecule} instances and the external molecular drivers. Leveraging this socket communication protocol, the EM solvers and molecular drivers can be initialized on separate computing nodes or even distinct HPC systems, thereby enabling large-scale light-matter simulations. For local simulations on a single computing node, the Python-based molecular drivers can also attach directly to the abstract \texttt{Molecule} instances, bypassing the socket communication layer; however, the C++ molecular driver (our modified LAMMPS code) functions exclusively in socket mode. The design of the socket communication layer is inspired by the open-source i-PI code for advanced MD simulations \cite{Litman2024}. 

    During the initialization phase, the molecular drivers configure themselves according to the time step $\Delta t$ defined by the EM solver. During time propagation, each abstract \texttt{Molecule} instance (indexed by $m$) calculates the regularized electric field vector $\widetilde{\vE}_m$ at its location and transmits $\widetilde{\vE}_m$ to the \texttt{SocketHub}. After gathering $\{\widetilde{\vE}_m\}$ from all \texttt{Molecule} instances, the \texttt{SocketHub} distributes each vector to its corresponding independent molecular driver. In turn, the driver returns the time derivative of the molecular dipole vector, which is used to construct the molecular current density for FDTD calculations. For facilitating normal-mode photonic simulations, the driver also returns the molecular dipole vector to the \texttt{SocketHub}.
        
    Since time-dependent molecular states are stored within the individual drivers, ensuring robustness is critical. After all, drivers may lose connection during simulations even within stable network environments. To address this issue, we implement several strategies: (i) The \texttt{SocketHub} returns dipole information to the EM solvers to advance to the next time step only after all drivers have successfully completed their calculations for the current step. (ii) If a molecular driver disconnects, the simulation pauses to await reconnection, while the hub preserves the field data for the interrupted step. (iii) Upon the launch of a new molecular driver to replace the failed one (which involves reading the latest internal molecular state from the disk), the driver receives the preserved field data. This ensures it rejoins the simulation and recomputes the interrupted step, restoring a state consistent with its peers. (iv) In all Python drivers, a two-phase commit model is employed for additional stability. At each time step, the drivers perform a trial step, calculating dipole information without immediately updating their internal state. Once the hub confirms receipt of messages from all drivers, it collects the data, and all drivers commit their new internal molecular states. This model ensures all molecular drivers to advance synchronously. (v) Beyond these internal safeguards, users are encouraged to frequently save checkpoints comprising the EM field and molecular states to facilitate simulation restarts.

    The three EM solvers and six molecular drivers currently available in \MaxwellLink are summarized in Tables \ref{table:em_solvers} and \ref{table:molecular_drivers}.  Users can view the \MaxwellLink documentation website for a comprehensive description of these solvers and drivers plus tutorials for launching various light-matter simulations \cite{MaxwellLinkDocument}. A sample Python input file for launching \MaxwellLink simulations is provided in Code Listing \ref{code:sample} for reference. It is also worth-noting that \MaxwellLink uses atomic units when communicating with molecular drivers and EM solvers; see Appendix A regarding units conversion between the native units in the MEEP FDTD code and atomic units. The boundary conditions of the EM field are taken care of by each EM solver. For example, the MEEP FDTD code supports absorbing, periodic, and metallic boundary conditions.

    The \MaxwellLink package is openly accessible on Github (\url{https://github.com/TaoELi/MaxwellLink}) and supports the standard Python package installer (\texttt{pip}).

    \begin{lstlisting}[language=Python, caption={Sample Python input for \MaxwellLink on propagating self-consistent EM-molecular simulations. Meep FDTD code is used as the EM solver, and the TCP/UNIX socket interface is applied for EM-molecular communication.}, label={code:sample}]
import maxwelllink as mxl
import maxwelllink.sockets as mxs
import meep as mp

host, port = mxs.get_available_host_port()
hub = mxl.SocketHub(host=host, port=port, timeout=10.0, latency=1e-4)

molecule = mxl.Molecule(
            hub=hub,
            center=mxl.Vector3(0, 0, 0),
            size=mxl.Vector3(1, 1, 1),
            sigma=0.1,
            dimensions=2,
        )

sim = mxl.MeepSimulation(
            hub=hub,
            molecules=[molecule], 
            time_units_fs=0.1,
            cell_size=mp.Vector3(8, 8, 0),
            boundary_layers=[mp.PML(3.0)],
            resolution=10,
        )

# Molecular drivers can run in terminals as:
# mxl_driver --model tls ...
# or using a Python wrapper of the above command
launch_driver(host, port) 

sim.run(steps=400)
    \end{lstlisting}
    
\section{Results} \label{sec:results}

    To demonstrate the capabilities and flexibility of the \MaxwellLink package, we  present four illustrative examples spanning from model-system to large-scale calculations. The simulation parameters are described in Appendix B.

     \begin{figure}
		\centering
		\includegraphics[width=1.0\linewidth]{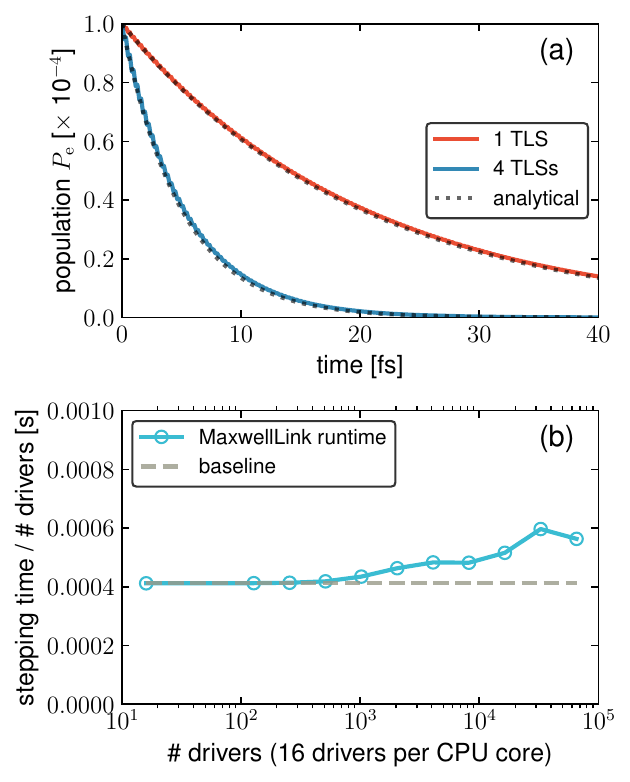}
		\caption{\textbf{Superradiance of $N$ TLSs in 2D vacuum.} (a) Time-resolved excited-state population dynamics where the TLSs are initialized in the same coherent state. Simulation results for $N=1$ (solid red) and $N=4$ (solid blue) TLSs are compared with the corresponding analytical spontaneous emission decay dynamics (black dotted). (b) Time cost for TCP socket communication per TLS driver. Up to $2^{16}$ independent TLS drivers, initialized across 32 computing nodes (4096 CPU cores), connect to the MEEP EM solver concurrently via TCP socket communication.
		}
		\label{fig:superradiance}
    \end{figure}

    \subsection{Superradiance in vacuum: Connecting to many drivers}

    A pivotal feature of \MaxwellLink is the employment of regularized electric fields $\widetilde{\vE}(t)$ to describe light-matter interactions. As discussed in Sec. \ref{sec:theory}, introducing regularized electric fields offers two primary advantages: (i) eliminating numerical singularities \cite{Lopata2009-1,Lopata2009-2} arising from the interaction between molecules and their self-emitted electric fields in FDTD; and (ii) significantly reducing the communication cost between the EM solver and molecular drivers.

    To assess the efficacy of regularized electric fields in \MaxwellLink, Fig. \ref{fig:superradiance} presents the radiative decay dynamics of $N$ TLSs collocated in 2D vacuum. Quantum-mechanically, when these TLSs are initialized in the same quantum state, their excited-state population $P_{\rm e}$ follows a uniform exponential decay:
    \begin{equation}\label{eq:population_decay_superradiance}
        P_{\rm e}(t) = P_{\rm e}(t=0) e^{-N \gamma_{\rm{QM}} t},
    \end{equation}
    where $P_{\rm e}(t=0)$ denotes the initial electronic excited-state population, and $\gamma_{\rm{QM}}$ represents the spontaneous emission rate for a single TLS in 2D vacuum \cite{Vargas2024}: $\gamma_{\rm{QM}} = \frac{|\mu_{\rm{ge}}|^2 \omega_{\rm{TLS}}^2}{2\hbar\epsilon_0 c^2}$.
    The $N$-scaling in the exponent of Eq. \eqref{eq:population_decay_superradiance} is attributed to Dicke superradiance \cite{Dicke1954}, wherein the spontaneous emission rate of the ensemble is enhanced linearly with respect to the number of molecules under the long-wave approximation. This enhancement stems from  the constructive interference of electric fields emitted by all other TLSs.
    
    Fig. \ref{fig:superradiance}a depicts the excited-state population dynamics of a single TLS (solid red) and four TLSs (solid blue) obtained from semiclassical \MaxwellLink simulations utilizing classical EM fields. Here, all TLSs start from the same initial coherent state $(c_{\rm g}, c_{\rm e}) = (\sqrt{0.9999}, 0.01)$. In these semiclassical simulations, the decay of the excited-state population is induced solely by the classical radiative self-interaction between the TLSs and their emitted electric fields, while quantum fluctuations of the EM field are excluded. As established in previous literature \cite{Jaynes1963,Li2018Spontaneous,Chen2018Spontaneous}, such semiclassical radiative decay dynamics reproduce full quantum results exactly in (and only in) the weak excitation limit. Consistent with this conclusion, our semiclassical decay dynamics agree with the analytical quantum dynamics [Eq. \eqref{eq:population_decay_superradiance}, dotted black lines], given the initial condition $|c_{\rm e}(t=0)|^2=10^{-4}$ within the weak excitation limit. This agreement confirms that the use of regularized electric fields successfully removes numerical singularities in the light-matter interaction.

    Fig. \ref{fig:superradiance}b further evaluates the \MaxwellLink stepping time by incorporating up to $N=2^{16}$ independent TLS drivers. These drivers, initialized on 32 computing nodes (utilizing 4096 CPU cores) on the Purdue Anvil HPC system \cite{Boerner2023}, connect to the MEEP EM solver via TCP socket communication. Under this connection mode, the expected \MaxwellLink stepping time can be decomposed into:
    \begin{equation}\label{eq:time_cost}
        \Delta t_{\rm{total}} = \Delta t_{\rm{EM}}  + \text{Max}[\Delta t_{\rm{driver}}] + \Delta t_{\rm{socket}},
    \end{equation}
    which comprises the stepping time for the EM solver ($\Delta t_{\rm{EM}}$), the slowest molecular driver ($\text{Max}[\Delta t_{\rm{driver}}]$), and the socket communication latency ($\Delta t_{\rm{socket}}$). In this 2D superradiance example, both $\Delta t_{\rm{EM}}$ and the internal computation time of the drivers are negligible due to the use of simplified model systems; consequently, as shown in Eq. \eqref{eq:time_cost}, the major contribution $\Delta t_{\rm{socket}}$ is expected to scale linearly with the number of TCP drivers connected to the EM solver in the ideal limit.
    
    Fig. \ref{fig:superradiance}b demonstrates that $\Delta t_{\rm{total}}/N$ remains stable at approximately $0.5$ ms for up to $N=2^{16}$ independent drivers connected concurrently to the EM solver. This suggests a uniform socket communication time per driver ($\sim 0.5$ ms) for inter-node communication via TCP sockets. This stable and low latency is achievable because the communication payload per driver is minimal (three floating-point numbers) due to the implementation of regularized electric fields. 
    
\begin{figure}
		\centering
		\includegraphics[width=1.\linewidth]{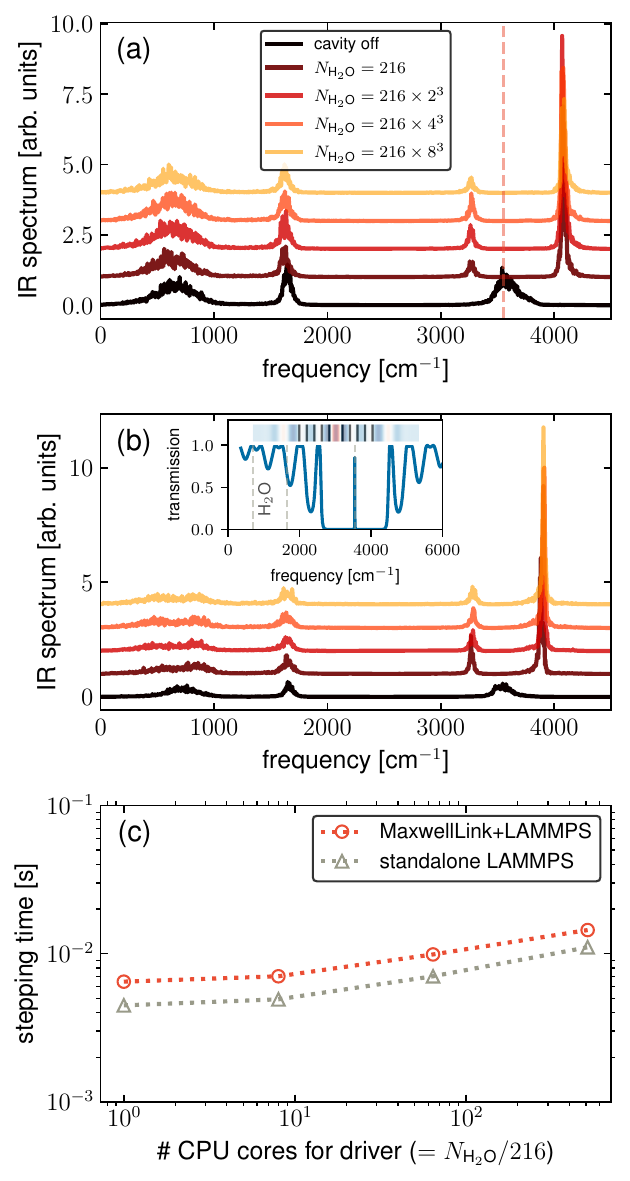}
		\caption{\textbf{Vibrational strong coupling of liquid water using \MaxwellLink.} (a) IR spectra of liquid water under single-mode CavMD, where a lossless cavity mode at $\omega_{\rm{c}}=3550  \ \text{cm}^{-1}$ (vertical dashed line) is coupled to the liquid water dipole moment along the $z$-axis. (b) Corresponding Maxwell-MD simulation results for liquid water confined in a 1D Bragg resonator. The inset displays a visualization of the 1D Bragg resonator (dielectric layers in the background and EM fields represented by color gradients) and the corresponding transmission spectrum. The three dashed vertical lines in the inset correspond to the three vibrational bands of liquid water. In the single-mode CavMD and Maxwell-MD simulations, a single LAMMPS driver, containing varying numbers of \ch{H2O} molecules, is coupled to the single-mode or MEEP FDTD EM solver, respectively. (c) \MaxwellLink stepping time versus the number of CPU cores used by the LAMMPS driver. The driver code utilizes MPI for parallel calculations of $N_{\ch{H2O}}$ \ch{H2O} molecules, with the number of CPU cores equal to $N_{\ch{H2O}}/216$.}
		\label{fig:vsc}
    \end{figure}

    Importantly, the demonstration with $2^{16}$ drivers does not imply that \MaxwellLink simulations of collective light-matter interactions are bounded by this number. Since each driver connected via TCP sockets can represent an ensemble of multiple real molecules (see Fig. \ref{fig:vsc}), the total molecular count can be significantly higher. In realistic scenarios, to optimize communication costs, it is computationally advantageous to connect the EM solver to a smaller set of molecular drivers (e.g., several thousand), each containing many realistic molecules and experiencing a unique EM environment. With this strategy, the explicitly simulated number of molecules in light-matter dynamics may potentially approach the macroscopic limit.

    \begin{figure}
		\centering
		\includegraphics[width=1.0\linewidth]{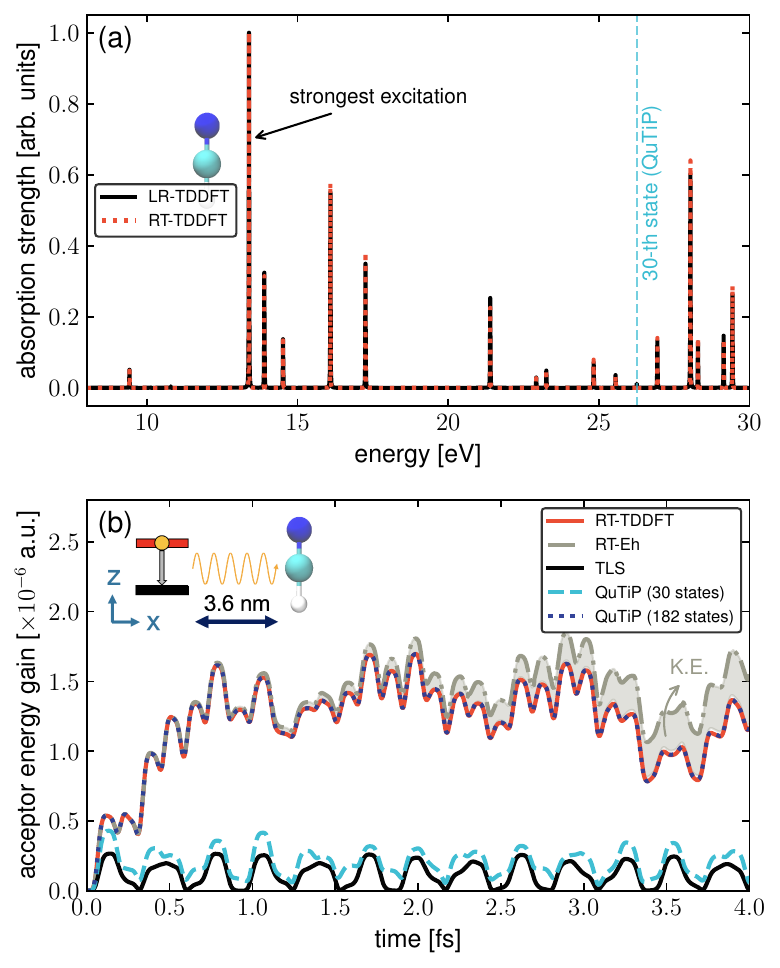}
		\caption{\textbf{Radiative energy transfer from a TLS donor to an HCN acceptor in 3D vacuum.} (a) Excited-state spectrum of the \ch{HCN} molecule at its optimized geometry, calculated at the B3LYP/cc-pVDZ level of theory. The in-house RT-TDDFT code within \MaxwellLink (dotted red) is compared with the LR-TDDFT calculations available in Psi4 (solid black). (b) Energy gain in the \ch{HCN} acceptor for an excited TLS donor transferring energy via classical electric fields (setup shown in the inset). The HCN acceptor is modeled using varying levels of theory: RT-TDDFT (solid red), RT-Ehrenfest dynamics (dash-dotted gray), a TLS containing only the strongest electronic transition in \ch{HCN} (solid black), and multi-level models using the QuTiP interface containing the lowest 30 (dashed cyan, with the frequency cutoff labeled in part a) or 182 (dotted blue) TDDFT singlet states. The kinetic energy contribution in the RT-Ehrenfest dynamics is highlighted by the shadowed gray region.}
		\label{fig:energy_transfer}
    \end{figure}

    \subsection{Resonance energy transfer: Heterogeneous simulations}

    The second important feature of \MaxwellLink is the capability of performing light-matter simulations with heterogeneous molecular drivers, as the modular implementation in \MaxwellLink avoids exposing molecular details directly to the EM solver. This feature enables multiscale light-matter simulations, where specific molecules are described by high-level theory while the remaining are treated using reduced models.

    Fig. \ref{fig:energy_transfer} showcases this heterogeneous simulation capability using the example of resonance energy transfer \cite{Li2018Tradeoff,Salam2018} from a TLS donor to an HCN molecule acceptor in 3D vacuum. The HCN molecule is described at the B3LYP/cc-pVDZ level \cite{Lee1988,Becke1988,Becke1998,Dunning1989}. With the optimized geometry aligned along the $z$-direction, as shown in Fig. \ref{fig:energy_transfer}a, the HCN excited-state spectrum calculated using the in-house RT-TDDFT code from \MaxwellLink (dotted red) agrees exactly with that obtained from linear-response TDDFT (LR-TDDFT, solid black) \cite{Li1986TDDFTMulti,Marques2006TDDFT,Butriy2007,Yang2018} via built-in Psi4 routines. Notably, the strongest electronic transition appears at $\omega_{\rm e} = 13.384$ eV, with the corresponding transition dipole moment $\mu_{\rm e}= 1.703$ a.u. oriented along the $z$-direction.
    
    Then, a TLS donor with this frequency $\omega_{\rm e} = 13.384$ eV and a larger transition dipole moment $\mu_{\rm e}'= 170.3$ a.u. is placed at a distance of 3.6 nm from the HCN molecule (Fig. \ref{fig:energy_transfer}b inset). When the TLS is initialized in a coherent state $(c_{\rm g}, c_{\rm e}) = (1/\sqrt{2}, 1/\sqrt{2})$ and the HCN molecule starts from the ground state, the TLS radiates classical electric fields, subsequently driving the HCN molecule through resonance energy transfer.

    Fig. \ref{fig:energy_transfer}b depicts the energy gain in the HCN molecule during the resonance energy transfer process. The energy gain described by RT-TDDFT (solid red) rises during the initial 1 fs and then reaches a plateau. When nuclear motion is allowed, the energy gain using RT-Ehrenfest dynamics (dash-dotted gray) matches the RT-TDDFT result exactly during the first 1 fs, whereas the subsequent nuclear motion in RT-Ehrenfest dynamics leads to a mild enhancement of the energy gain at later times; see the shadowed gray region for the corresponding kinetic energy contribution.

    When a TLS is used to represent this HCN acceptor with only the strongest TDDFT electronic transition included (solid black), the corresponding energy gain is significantly reduced. Going beyond the TLS approximation, if the HCN acceptor is modeled by a multi-level model Hamiltonian using the QuTiP interface, including the 30 lowest TDDFT singlet states (dashed cyan) yields dynamics similar to those of the TLS acceptor, with the highest included frequency in the QuTiP model labeled as the vertical cyan line in Fig. \ref{fig:energy_transfer}a.  By contrast, including all 182 TDDFT singlet states (dotted blue) recovers the RT-TDDFT limit (solid red).
    
    Overall, this simulation not only cross-validates various molecular drivers, but also demonstrates the flexibility of \MaxwellLink for multiscale light-matter simulations. In particular, switching between different levels of theory for describing the HCN acceptor requires changing only a single line in the input file.

    \subsection{Vibrational strong coupling: From single-mode CavMD to Maxwell-MD}

    Apart from supporting multiscale molecular simulations, \MaxwellLink also enables users to flexibly vary the level of theory for the EM solver in light-matter simulations. This feature provides a convenient platform for examining the approximations used for modeling the EM field.
    
    As an illustrative example, Fig. \ref{fig:vsc} compares vibrational strong coupling for liquid water confined in a single-mode cavity versus a realistic 1D Bragg resonator. Building upon previous CavMD simulations of liquid water under vibrational strong coupling \cite{Li2020Water}, Fig. \ref{fig:vsc}a reports the IR spectra of liquid water in free space (bottom black line) versus those coupled to a single-mode cavity (colored lines). This classical single-mode cavity is polarized along the $z$-direction, with a frequency $\omega_{\rm c} = 3550$ cm$^{-1}$ (vertical orange line) set at resonance with the \ch{O-H} stretch band of liquid water, utilizing the q-TIP4P/F force field via the LAMMPS driver. When this cavity mode is coupled to $N_{\ch{H2O}}=216$ molecules, an effective coupling strength of $\varepsilon = 4\times 10^{-4}$ a.u. [c.f. Eq. \eqref{eq:varepsilon_cavmd}] yields a Rabi splitting of approximately 800 cm$^{-1}$ (brown line). As the molecular number increases by factors of $M=2^3$, $4^3$, and $8^3$ (corresponding to the orange through yellow lines, respectively), the Rabi splitting remains constant, provided the light-matter coupling scales as $\varepsilon' = \varepsilon / \sqrt{M}$.

    Moving beyond the single-mode CavMD implemented in \MaxwellLink using the single-mode EM solver, employing the MEEP FDTD solver enables coupled Maxwell-MD simulations of vibrational strong coupling. For simplicity, we replace the single-mode cavity here with a 1D Bragg resonator \cite{Skolnick1998}. As shown in the inset of Fig. \ref{fig:vsc}b, this 1D Bragg resonator  consists of two mirrors separated by $\lambda/2$, with each mirror composed of five periodic dielectric layers spaced by $\lambda/4$. With $\lambda= 2.818\ \mu$m, this Bragg resonator supports a cavity mode at frequency 3550 cm$^{-1}$, as indicated by the cavity transmission spectrum in the inset.

    Fig. \ref{fig:vsc}b plots the corresponding IR spectra of liquid water when the molecular system is placed at the center of the 1D Bragg resonator. With $N_{\ch{H2O}}=216$ molecules explicitly included (brown line), we achieve strong coupling for the water \ch{O-H} stretch mode by artificially enhancing the electric current density of the water system by a factor of $\gamma=10^5$ [c.f. Eq. \eqref{eq:Polarization_mol}]. Additionally, due to the inclusion of a realistic cavity geometry, the librational band of liquid water around 700 cm$^{-1}$ also splits into a pair of polariton peaks. This occurs because the librational band (indicated by the left vertical dashed line in the inset) coincides with a low-frequency mode of the Bragg resonator.
    
    Similar to the single-mode simulation case, when the molecular number increases by factors of $M=2^3$, $4^3$, and $8^3$ (from orange to yellow lines, respectively), the Rabi splitting of approximately 600 cm$^{-1}$ remains constant when the rescaling factor is reduced to $\gamma' = \gamma/\sqrt{M}$. This trend suggests that vibrational strong coupling of liquid water in a 1D Bragg resonator requires approximately $2\times 10^{12}$ \ch{H2O} molecules (where $M=10^{10}$ and $\gamma' = 1$). We note that the Maxwell-MD simulation directly yields this molecular number without estimating the effective cavity volume $\mathcal{V}$, whereas simplified approaches, such as single-mode CavMD (which relies on the effective coupling strength $\varepsilon$, depending on the effective cavity volume $\mathcal{V}$), cannot directly provide this quantity.

    Technically speaking, the simulations in Fig. \ref{fig:vsc} involve the connection between an EM solver and a single LAMMPS molecular driver via TCP sockets. For simulations with enlarged molecular systems, Fig. \ref{fig:vsc}c reports the \MaxwellLink stepping time as the number of CPU cores used for the MPI parallelism of the LAMMPS code, as the CPU core number increases proportionally to the simulated molecular number. Compared to the standalone LAMMPS simulation of liquid water (gray triangles), the modest increase in stepping time for \MaxwellLink coupled with LAMMPS demonstrates the efficiency of our implementation.

    \begin{figure*}
		\centering
		\includegraphics[width=0.95\linewidth]{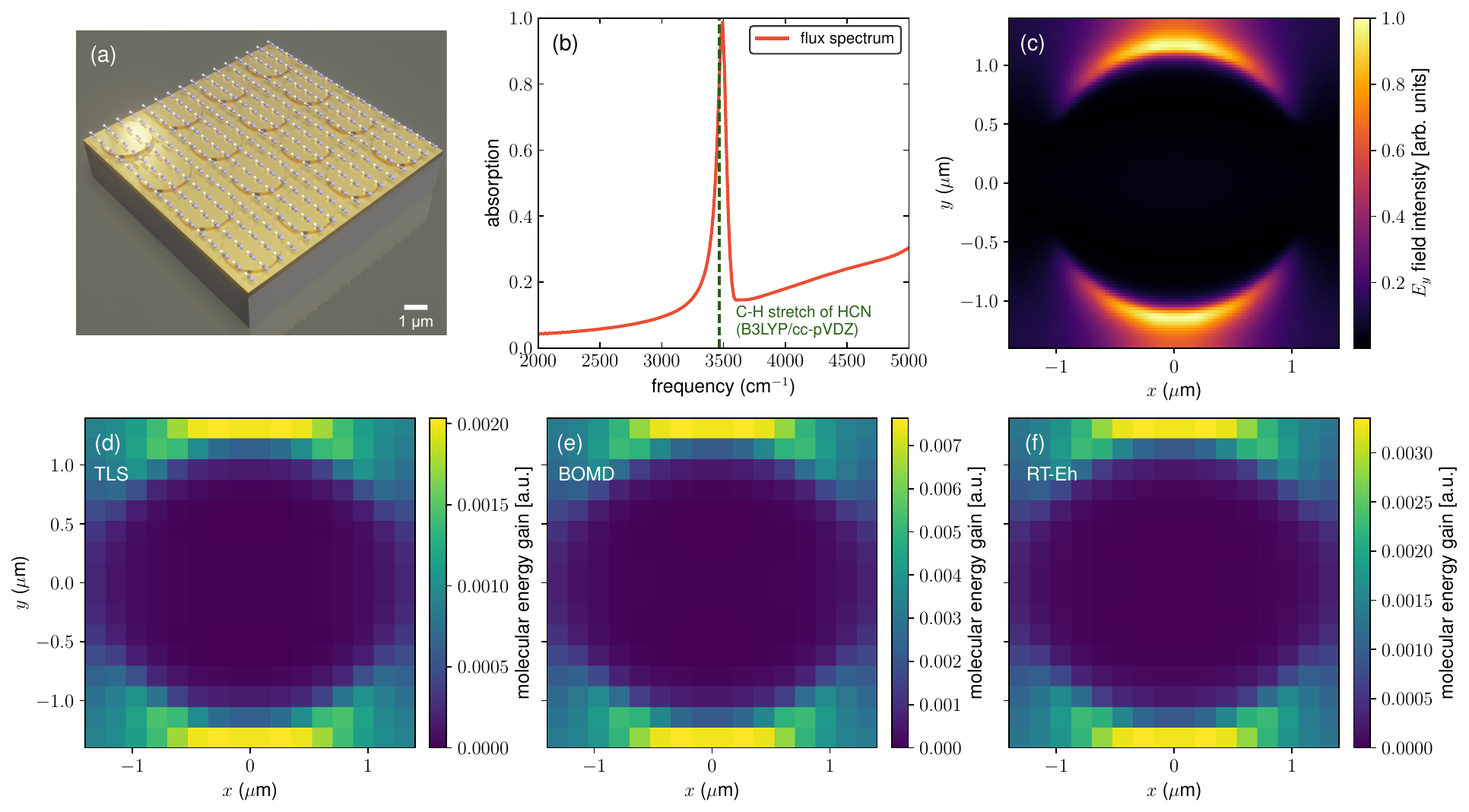}
		\caption{\textbf{Vibrational heating of HCN molecules on top of 3D plasmonic metamaterials.} (a) The simulation setup contains a square lattice of HCN molecules (oriented along the $y$-direction) above a 3D plasmonic metamaterial consisting of a square lattice of cylindrical Pt rods on a semi-infinite Si substrate. A single unit cell containing one Pt rod and 256 HCN molecules is simulated, with periodic boundary conditions applied along the $xy$-plane and absorbing boundary conditions along the $z$-direction. (b) Absorption spectrum of the plasmonic metamaterial in the absence of HCN molecules. This geometry supports a surface plasmonic mode near resonance with the \ch{C-H} stretch mode of HCN at $\omega_{\rm v}=3466$ cm$^{-1}$ (dashed green). (c) Electric field intensity distribution in a unit cell following a $y$-polarized Gaussian pulse excitation of the plasmonic mode in the absence of molecules. (d)-(f) Real-space energy gain distribution of individual HCN molecules under the same Gaussian pulse excitation of the hybrid plasmonic-molecular system. The molecules are simulated using (d) TLSs, (e) first-principles Born--Oppenheimer MD, and (f) RT-Ehrenfest dynamics at the B3LYP/cc-pVDZ level of theory.}
		\label{fig:plasmon_heating}
    \end{figure*}

    \subsection{Vibrational heating of molecules near plasmonic metamaterials}

    As a final demonstrative example, we apply \MaxwellLink to explore how a realistic 3D IR plasmonic metamaterial induces spatially inhomogeneous vibrational heating within a molecular ensemble.

    As illustrated in Fig. \ref{fig:plasmon_heating}a, this plasmonic metamaterial consists of a square lattice of cylindrical platinum (Pt) rods supported by a semi-infinite silicon (Si) substrate \cite{Inoue2013}. In the absence of adsorbed molecules, FDTD simulations reveal that the metamaterial supports a surface plasmonic mode at frequency $\omega_{\rm p} = 3492$ cm$^{-1}$ (Fig. \ref{fig:plasmon_heating}b) when excited by a plane-wave Gaussian pulse at normal incidence propagating along the $z$-axis. Because the incident Gaussian pulse is polarized along the $y$-direction, the electric field distribution of this plasmonic mode (Fig. \ref{fig:plasmon_heating}c) exhibits strong enhancement within the gaps between neighboring cylindrical Pt rods along the $y$-axis. Leveraging the periodicity of the metamaterial in the $xy$-plane, a single unit cell comprising a Pt rod and the Si substrate is simulated via FDTD. The simulation domain spans $2.8\times2.8\times 16.5$ $\mu$m$^3$, corresponding to a total of $8.28\times 10^6$ grid points. To handle this computationally demanding calculation, MPI parallel acceleration is employed for the MEEP FDTD engine using 128 CPU cores.

    Subsequently, a square lattice of $N_{\ch{HCN}} =256$ \ch{HCN} molecules is positioned within this unit cell in the $xy$-plane, 0.125 $\mu$m above the top surface of the Pt rod. The \ch{HCN} molecules are described at the B3LYP/cc-pVDZ level of theory and possess a high-frequency \ch{C-H} stretch mode at $\omega_{\rm{v}}=3466$ cm$^{-1}$ (Fig. \ref{fig:plasmon_heating}b, dashed green line). Because this vibrational mode is nearly resonant with the plasmonic frequency, we apply the same Gaussian pulse to excite the hybrid plasmonic-molecular system, propagating the molecular dynamics  using 256 independent molecular drivers in \MaxwellLink. Here, the intermolecular interactions between the \ch{HCN} molecules are propagated solely by the classical EM field. This treatment should be valid in this case, as the nearest-neighbor spacing between the \ch{HCN} molecules is 0.175 $\mu$m.
    
    Due to the flexibility of \MaxwellLink, the HCN molecules are propagated using TLSs, first-principles Born--Oppenheimer MD, or RT-Ehrenfest dynamics, as shown in Figs. \ref{fig:plasmon_heating}d-f. However, regardless of the level of theory for describing the molecules, during pulse excitation, molecules located within the gaps along the $y$-direction exhibit greater energy gain than the remaining molecules.   This result is consistent with the electric field intensity distribution obtained from the FDTD simulation in the absence of \ch{HCN} molecules (Fig. \ref{fig:plasmon_heating}c). Note that the difference of absolute numbers in the energy gain of the three methods arises from the approximations made in defining light-matter coupling (Appendix B). 
    
    This example highlights the capability of \MaxwellLink to perform large-scale light-matter simulations involving realistic 3D plasmonic geometries interacting with a large collection of first-principles molecules. Beyond this simulation, condensed-phase molecules coupled to 3D plasmonic geometries may also be simulated within the framework of \MaxwellLink using the following strategy: Each driver containing a small condensed-phase molecular simulation cell accounts for the short-range intermolecular interactions, leaving the long-range interactions between different molecular drivers propagated by the classical EM field via \MaxwellLink.

   \section{Conclusion and Outlook}\label{sec:conclusion}

    In summary, we have introduced the \MaxwellLink package, which provides a flexible and unified framework for self-consistent light-matter simulations. In contrast to existing light-matter simulation codes, \MaxwellLink offers a uniform Python interface for flexibly switching between levels of theory for both the EM solver and molecules, ranging from model systems to large-scale realistic calculations. Additionally, through a TCP/UNIX socket interface and an MPI-aware implementation, it enables the simultaneous parallel acceleration of the EM solver and molecular drivers across different computational nodes or even separate HPC systems --- a feature that is particularly appealing for large-scale light-matter simulations. \MaxwellLink is also accompanied by a detailed documentation website and tutorials \cite{MaxwellLinkDocument}, facilitating the adoption of the package by newcomers and students.

    The examples presented in this manuscript showcase the potential of \MaxwellLink as a research tool in fields such as quantum optics, plasmonics, polaritonics, and spectroscopy. More extensive applications of \MaxwellLink will be reported in future work.

    Looking forward, additional EM solvers and molecular drivers, particularly advanced electronic structure packages or GPU-accelerated EM solvers, may be integrated into \MaxwellLink to enable more efficient, large-scale production calculations. This package can also serve as a versatile platform for method development aimed at addressing fundamental limitations of the current self-consistent light-matter scheme, such as the classical approximation of the EM field and the restriction to purely electric field interactions. For example, it is known that using classical EM fields alone cannot describe spontaneous emission and resonance energy transfer beyond the weak excitation limit \cite{Li2018Spontaneous,Chen2018Spontaneous,Li2018Tradeoff} --- resolving this issue in \MaxwellLink is an exciting direction. Overall, the open-source \MaxwellLink framework has the potential to transform self-consistent light-matter simulations from a niche computational tool used by a handful of research groups into a widely accessible resource for the broader scientific community.

    \section*{Acknowledgement}
    This material is based upon work supported by the U.S. National Science Foundation under Grant No. CHE-2502758. This work used the Anvil HPC at Purdue University through allocation CHE250091 from the Advanced Cyberinfrastructure Coordination Ecosystem: Services \& Support (ACCESS) program, which is supported by U.S. National Science Foundation grants \#2138259, \#2138286, \#2138307, \#2137603, and \#2138296.

    \section*{Data and Software Availability}
      The simulation examples presented in this manuscript were performed using \MaxwellLink v0.2, the initial public release of the code available on GitHub: \url{https://github.com/TaoELi/MaxwellLink}. The input scripts and post-processing files for these examples are archived in a separate GitHub repository (\url{https://github.com/TaoELi/maxwelllink_examples}).
    
    \appendix

    \section*{Appendix A: Units conversion between \MaxwellLink and MEEP}\label{app:units}

    In \MaxwellLink, atomic units are employed for communication between the abstract \texttt{Molecule} instances and the molecular drivers. However, because FDTD engines typically utilize different unit systems, the conversion between the FDTD native units and atomic units requires further clarification.
    
    For example, MEEP operates in a dimensionless unit system where $c = \epsilon_0 = \mu_0 = 1$. By additionally setting $\hbar=1$ and defining the FDTD time unit as $\tau_{\text{fs}}$ (in femtoseconds), the unit system can be fully determined. Consequently, the time step in MEEP is converted to atomic units as:
    \begin{equation}\label{eq:units_time_au_meep}
        \Delta t_{\rm{a.u.}} = \Delta t_{\rm{MEEP}} \times \tau_{\rm{fs}} \times 41.341373 .
    \end{equation}
    In Eq. \eqref{eq:units_time_au_meep}, the constant 41.341373 converts 1 fs to atomic units. During the initial communication handshake, \MaxwellLink instructs the molecular drivers to use $\Delta t_{\rm{a.u.}}$ for propagating molecular dynamics. 

    Upon evaluation of the regularized electric field vector $\widetilde{\vE}$, \MaxwellLink converts the electric field from MEEP units to atomic units via:
    \begin{equation}\label{eq:E_conversion}
        \widetilde{\vE}_{\rm{a.u.}} = \frac{1.292954\times 10^{-6} }{\tau_{\rm{fs}}^2}  \widetilde{\vE}_{\rm{MEEP}} .
    \end{equation}
    This conversion relation is derived as follows. One atomic unit of electric field is defined as $|e|/4\pi\epsilon_0 r_{\rm{Bohr}}^2$, where $e$ and $r_{\rm{Bohr}}$ denote the elementary charge and Bohr radius, respectively. In MEEP units, where $\epsilon_0 = c = \hbar = 1$, the dimensionless fine structure constant $\alpha = e^2/4\pi \epsilon_0 \hbar c$ yields $|e| = \sqrt{4\pi \alpha}$. Given the MEEP time unit $[T] = \tau_{\rm{fs}} \times 1$ fs, the length scale is $r_{\rm{Bohr}}$ = $1.765101\times 10^{-4} c$ fs = $1.765101\times 10^{-4} c / \tau_{\rm{fs}}$ in MEEP units. Hence, 1 a.u. of electric field corresponds to $\sqrt{4\pi \alpha} / [4\pi (1.765101\times 10^{-4} / \tau_{\rm{fs}})^2]  = 7.734616\times 10^5\times \tau_{\rm{fs}}^2$ in MEEP units, which represents the inverse of Eq. \eqref{eq:E_conversion}.
   
    Additionally, \MaxwellLink also converts the dipole time derivative returned by the driver from atomic units to MEEP units:
    \begin{equation}\label{eq:mudot_conversion}
        \dot{\vmu}_{\rm{MEEP}} = 2.209800 \times 10^{-3} \times \dot{\vmu}_{\rm{a.u.}} .
    \end{equation}
    This conversion is determined by preserving the dimensionless quantity $t^2 \widetilde{\vE}\cdot \dot{\vmu} /\hbar$. By setting $t=1$ fs, the equality becomes $\tau_{\rm {fs}}^{-2} \widetilde{\vE}_{\rm{MEEP}} \cdot \dot{\vmu}_{\rm{MEEP}}  =  41.3413745758^2 \times \widetilde{\vE}_{\rm{a.u.}} \cdot \dot{\vmu}_{\rm{a.u.}} $.  When further combined with Eq. \eqref{eq:E_conversion}, this equality yields the conversion factor in Eq. \eqref{eq:mudot_conversion}.
    
    \section*{Appendix B: Simulation Details}\label{app:simu_details}

    To facilitate the reproduction of the reported examples,  we provide the FDTD parameters in Table \ref{table:parameters}. Note that the geometric center of the FDTD simulation cell was set as the origin of the coordinate system, and the spatial kernel function of each \texttt{Molecule} instance assumed a normalized Gaussian distribution with a width parameter $\sigma=\Delta x$, matching the FDTD spatial resolution. For all simulations, the TCP socket interface was applied for the communication between EM solvers and molecular drivers. The other important simulation details are  outlined below.

    \begin{table*}
    \caption{FDTD parameters for the four examples in Sec. \ref{sec:results}.
    }
    \label{table:parameters}
    \begin{tabular}{lcccccr}
    \hline
    \hline
    Example & \ \ Cell volume [$\mu$m$^3$] \ \ & PML thickness [nm] \ \  & $\Delta x$ [nm] & \ \ $\Delta t$ [fs]  &  \ \ \texttt{Molecule} location(s) [nm] \\
    \hline
    Superradiance & \ \ $0.24\times 0.24\times 0$ \ \ & 90 \ \  & 3 & \ \ 5$\times 10^{-3}$   & \ \ (0, 0, 0) \\
    Energy transfer & \ \ $0.048\times 0.048\times 0.048$ \ \ & 18 \ \  & 0.6 & \ \ $10^{-3}$   & \ \ ($\pm 1.8$, 0, 0) \\
    Strong coupling & \ \ $23.3\times 0\times 0$  \ \ & 5.64  $\times 10^{3}$ \ \  & 2.4 $\times 10^2$ & \ \ 0.4   &  \ \ (0, 0, 0) \\
    Plasmonic heating & \ \ $2.8\times 2.8\times 16.5$ \ \ & 5 $\times 10^3$ ($z$ only) \ \  & 25 & \ \ 4.2 $\times 10^{-2}$   & \ \ 2D square lattice \\
    \hline
    \hline
    \end{tabular}
    \end{table*}

    \subsection*{Superradiance} 
    
    All $N$ TLSs were positioned at the center of the 2D FDTD simulation cell within the $xy$-plane. Each TLS, characterized by a frequency $\omega_{\rm{TLS}} = 0.242$ a.u. and a transition dipole moment $\mu_{12} = 187$ a.u. aligned along the $z$-axis, was initialized in an identical coherent state $(c_{\rm g}, c_{\rm e}) = (\sqrt{0.9999}, 0.01)$. For benchmarks utilizing up to $N = 2^{16}$ TLS drivers, the transition dipole moment of individual TLSs was rescaled to $\mu_{12} = 187/\sqrt{N}$. This rescaling ensured that the radiative decay lifetime remained constant independent of the total number of TLSs, a condition crucial for accurately comparing the \MaxwellLink stepping time as a function of the connected drivers.

    \subsection*{Energy transfer} 
    
    3D FDTD calculations were performed in vacuum. The TLS donor frequency was set to $\omega_{\rm e} = 13.384$ eV, with a transition dipole moment of $\mu_{\rm e}'= 170.3$ a.u. aligned along the $z$-axis. The initial quantum state was prepared as the superposition $(c_{\rm g}, c_{\rm e}) = (\sqrt{1/2}, \sqrt{1/2})$. For the HCN acceptor, the B3LYP functional \cite{Lee1988,Becke1988,Becke1998} and the cc-pVDZ basis set \cite{Dunning1989} were employed for both RT-TDDFT and RT-Ehrenfest calculations. The HCN acceptor was initialized in its optimized ground-state geometry and oriented along the $z$-axis, separated from the TLS donor by a distance of 3.6 nm along the $x$-axis. For the RT-Ehrenfest dynamics, the Li--Tully--Schlegel--Frisch integration scheme (the nuclear-position-coupled midpoint Fock integrator) \cite{Li2005Eh} was implemented, augmented by an additional predictor-corrector procedure \cite{DeSantis2020} to improve accuracy, following a recent implementation \cite{Li2023eBO}. The electronic propagation time step was set identical to the FDTD $\Delta t$, while nuclear gradients were evaluated every 100 electronic propagation steps. When modeling the HCN acceptor as a TLS, the strongest TDDFT electronic transition ($\omega_{\rm e} = 13.384$ eV, $\mu_{\rm e}= 1.703$ a.u., oriented along the $z$-axis) was selected. When representing the HCN acceptor with a multilevel model Hamiltonian, the lowest 30 or 182 TDDFT singlet states (including the ground state) were included.

    \subsection*{Strong coupling} 
    
    The 1D Bragg resonator  comprised two mirrors separated by a distance of $\lambda/2$, where each mirror consisted of five periodic dielectric layers spaced by $\lambda/4$. Each dielectric layer, characterized by a refractive index of $n=2$, possessed a width of $\lambda/8$. Given $\lambda= 2.818\ \mu$m, this Bragg resonator supports a cavity mode at a frequency of 3550 cm$^{-1}$. 
    
    For simulations involving $N_{\ch{H2O}}=216$ water molecules, the initial configuration was adopted from a thermally equilibrated geometry at 300 K, as described in previous work \cite{Li2020Water} utilizing the q-TIP4P/F force field \cite{Habershon2009ZPE}. A single 100-ps NVE trajectory was computed to evaluate the IR spectrum. The spectrum was subsequently obtained by computing the Fourier transform of the dipole autocorrelation function of the liquid water system \cite{Li2020Water}. 
    
    For enlarged water systems, the initial geometry was generated by replicating the $N_{\ch{H2O}}=216$ water configuration, followed by a 20-ps NVT equilibration to relax the replicated system. All water simulations were conducted using the LAMMPS driver.

    \subsection*{Plasmonic heating} 
    
    The Pt/Si plasmonic metamaterial model was constructed to reproduce the FDTD simulations described in Ref. \citenum{Inoue2013}. Specifically, the cylindrical Pt rod possessed a radius of $r=1.11$ $\mu$m and a height of $0.2$ $\mu$m.  This rod was placed on top of a uniform Pt layer with a thickness of $0.3$ $\mu$m. Beneath the Pt layer, a uniform Si substrate with a thickness of 2.0 $\mu$m was employed. Perfectly matched layers (PML) with a thickness of 5.0 $\mu$m were applied at the top and bottom boundaries along the $z$-axis to implement absorbing boundary conditions. Periodic boundary conditions were imposed in the $xy$-plane; see also Ref. \cite{Simpetus} for detailed parameters.
    
    A total of 256 HCN molecules were distributed on a 2D square lattice in the $xy$-plane, positioned 0.125 $\mu$m above the top surface of the Pt rod. All HCN molecules were oriented along the $y$-axis. Each HCN molecule was initialized in the electronic ground state near its optimized geometry. The molecular electronic structure was described at the B3LYP/cc-pVDZ level of theory. 
    
    For Born--Oppenheimer MD simulations, the ASE driver was employed with fixed partial charges [cf. Eq. \eqref{eq:dipole_partial_charge}] assigned as $Q_{\rm{C}}=-0.247$, $Q_{\rm{N}}=-0.003$, and $Q_{\rm{H}}=0.250$, a major approximation which may not accurately evaluate light-matter coupling. For RT-Ehrenfest dynamics simulations, the nuclear time step was set equal to the FDTD $\Delta t$, while the electronic degrees of freedom were propagated for 36 substeps per nuclear step. In cases where the HCN molecules were modeled as TLSs, the transition frequency was set to $\omega_{\rm v} = 3466$ cm$^{-1}$, matching the high-frequency \ch{C-H} vibrational mode of \ch{HCN}. The transition dipole moment of each TLS was set to $\mu_{12} = 0.15$ a.u. along the $y$-axis. 
    
    A plane-wave Gaussian pulse, centered at frequency of $3.5$ $\mu$m$^{-1}$ with a width of 0.3 $\mu$m$^{-1}$, was applied to excite the system at normal incidence along the $z$-axis. The Gaussian pulse was polarized along the $y$-axis, with the amplitude set to $5.8\times 10^{-3}$ a.u. ($5\times 10^{4}$ in native MEEP units). Each simulation was propagated for a duration of 200 fs. The energy gain of each molecule was computed by taking the average energy during the 200-fs trajectory.
    

\begin{thebibliography}{71}%
\makeatletter
\providecommand \@ifxundefined [1]{%
 \@ifx{#1\undefined}
}%
\providecommand \@ifnum [1]{%
 \ifnum #1\expandafter \@firstoftwo
 \else \expandafter \@secondoftwo
 \fi
}%
\providecommand \@ifx [1]{%
 \ifx #1\expandafter \@firstoftwo
 \else \expandafter \@secondoftwo
 \fi
}%
\providecommand \natexlab [1]{#1}%
\providecommand \enquote  [1]{``#1''}%
\providecommand \bibnamefont  [1]{#1}%
\providecommand \bibfnamefont [1]{#1}%
\providecommand \citenamefont [1]{#1}%
\providecommand \href@noop [0]{\@secondoftwo}%
\providecommand \href [0]{\begingroup \@sanitize@url \@href}%
\providecommand \@href[1]{\@@startlink{#1}\@@href}%
\providecommand \@@href[1]{\endgroup#1\@@endlink}%
\providecommand \@sanitize@url [0]{\catcode `\\12\catcode `\$12\catcode `\&12\catcode `\#12\catcode `\^12\catcode `\_12\catcode `\%12\relax}%
\providecommand \@@startlink[1]{}%
\providecommand \@@endlink[0]{}%
\providecommand \url  [0]{\begingroup\@sanitize@url \@url }%
\providecommand \@url [1]{\endgroup\@href {#1}{\urlprefix }}%
\providecommand \urlprefix  [0]{URL }%
\providecommand \Eprint [0]{\href }%
\providecommand \doibase [0]{https://doi.org/}%
\providecommand \selectlanguage [0]{\@gobble}%
\providecommand \bibinfo  [0]{\@secondoftwo}%
\providecommand \bibfield  [0]{\@secondoftwo}%
\providecommand \translation [1]{[#1]}%
\providecommand \BibitemOpen [0]{}%
\providecommand \bibitemStop [0]{}%
\providecommand \bibitemNoStop [0]{.\EOS\space}%
\providecommand \EOS [0]{\spacefactor3000\relax}%
\providecommand \BibitemShut  [1]{\csname bibitem#1\endcsname}%
\let\auto@bib@innerbib\@empty
\bibitem [{\citenamefont {Odom}\ and\ \citenamefont {Schatz}(2011)}]{Odom2011}%
  \BibitemOpen
  \bibfield  {author} {\bibinfo {author} {\bibfnamefont {T.~W.}\ \bibnamefont {Odom}}\ and\ \bibinfo {author} {\bibfnamefont {G.~C.}\ \bibnamefont {Schatz}},\ }\bibfield  {title} {\bibinfo {title} {{Introduction to Plasmonics}},\ }\href {https://doi.org/10.1021/cr2001349} {\bibfield  {journal} {\bibinfo  {journal} {Chem. Rev.}\ }\textbf {\bibinfo {volume} {111}},\ \bibinfo {pages} {3667} (\bibinfo {year} {2011})}\BibitemShut {NoStop}%
\bibitem [{\citenamefont {Aroca}(2013)}]{Aroca2013}%
  \BibitemOpen
  \bibfield  {author} {\bibinfo {author} {\bibfnamefont {R.~F.}\ \bibnamefont {Aroca}},\ }\bibfield  {title} {\bibinfo {title} {{Plasmon Enhanced Spectroscopy}},\ }\href {https://doi.org/10.1039/c3cp44103b} {\bibfield  {journal} {\bibinfo  {journal} {Phys. Chem. Chem. Phys.}\ }\textbf {\bibinfo {volume} {15}},\ \bibinfo {pages} {5355} (\bibinfo {year} {2013})}\BibitemShut {NoStop}%
\bibitem [{\citenamefont {Wang}\ \emph {et~al.}(2020)\citenamefont {Wang}, \citenamefont {Huang}, \citenamefont {Hu}, \citenamefont {Yan},\ and\ \citenamefont {Ren}}]{Wang2020NatRevPhys}%
  \BibitemOpen
  \bibfield  {author} {\bibinfo {author} {\bibfnamefont {X.}~\bibnamefont {Wang}}, \bibinfo {author} {\bibfnamefont {S.-C.}\ \bibnamefont {Huang}}, \bibinfo {author} {\bibfnamefont {S.}~\bibnamefont {Hu}}, \bibinfo {author} {\bibfnamefont {S.}~\bibnamefont {Yan}},\ and\ \bibinfo {author} {\bibfnamefont {B.}~\bibnamefont {Ren}},\ }\bibfield  {title} {\bibinfo {title} {{Fundamental Understanding and Applications of Plasmon-Enhanced Raman Spectroscopy}},\ }\href {https://doi.org/10.1038/s42254-020-0171-y} {\bibfield  {journal} {\bibinfo  {journal} {Nat. Rev. Phys.}\ }\textbf {\bibinfo {volume} {2}},\ \bibinfo {pages} {253} (\bibinfo {year} {2020})}\BibitemShut {NoStop}%
\bibitem [{\citenamefont {Ribeiro}\ \emph {et~al.}(2018)\citenamefont {Ribeiro}, \citenamefont {Mart{\'{i}}nez-Mart{\'{i}}nez}, \citenamefont {Du}, \citenamefont {Campos-Gonzalez-Angulo},\ and\ \citenamefont {Yuen-Zhou}}]{Ribeiro2018}%
  \BibitemOpen
  \bibfield  {author} {\bibinfo {author} {\bibfnamefont {R.~F.}\ \bibnamefont {Ribeiro}}, \bibinfo {author} {\bibfnamefont {L.~A.}\ \bibnamefont {Mart{\'{i}}nez-Mart{\'{i}}nez}}, \bibinfo {author} {\bibfnamefont {M.}~\bibnamefont {Du}}, \bibinfo {author} {\bibfnamefont {J.}~\bibnamefont {Campos-Gonzalez-Angulo}},\ and\ \bibinfo {author} {\bibfnamefont {J.}~\bibnamefont {Yuen-Zhou}},\ }\bibfield  {title} {\bibinfo {title} {{Polariton Chemistry: Controlling Molecular Dynamics with Optical Cavities}},\ }\href {https://doi.org/10.1039/C8SC01043A} {\bibfield  {journal} {\bibinfo  {journal} {Chem. Sci.}\ }\textbf {\bibinfo {volume} {9}},\ \bibinfo {pages} {6325} (\bibinfo {year} {2018})}\BibitemShut {NoStop}%
\bibitem [{\citenamefont {Flick}\ and\ \citenamefont {Narang}(2018)}]{Flick2018}%
  \BibitemOpen
  \bibfield  {author} {\bibinfo {author} {\bibfnamefont {J.}~\bibnamefont {Flick}}\ and\ \bibinfo {author} {\bibfnamefont {P.}~\bibnamefont {Narang}},\ }\bibfield  {title} {\bibinfo {title} {{Cavity-Correlated Electron-Nuclear Dynamics from First Principles}},\ }\href {https://doi.org/10.1103/PhysRevLett.121.113002} {\bibfield  {journal} {\bibinfo  {journal} {Phys. Rev. Lett.}\ }\textbf {\bibinfo {volume} {121}},\ \bibinfo {pages} {113002} (\bibinfo {year} {2018})}\BibitemShut {NoStop}%
\bibitem [{\citenamefont {Herrera}\ and\ \citenamefont {Owrutsky}(2020)}]{Herrera2019}%
  \BibitemOpen
  \bibfield  {author} {\bibinfo {author} {\bibfnamefont {F.}~\bibnamefont {Herrera}}\ and\ \bibinfo {author} {\bibfnamefont {J.}~\bibnamefont {Owrutsky}},\ }\bibfield  {title} {\bibinfo {title} {{Molecular Polaritons for Controlling Chemistry with Quantum Optics}},\ }\href {https://doi.org/10.1063/1.5136320} {\bibfield  {journal} {\bibinfo  {journal} {J. Chem. Phys.}\ }\textbf {\bibinfo {volume} {152}},\ \bibinfo {pages} {100902} (\bibinfo {year} {2020})}\BibitemShut {NoStop}%
\bibitem [{\citenamefont {Li}\ \emph {et~al.}(2022)\citenamefont {Li}, \citenamefont {Cui}, \citenamefont {Subotnik},\ and\ \citenamefont {Nitzan}}]{Li2022Review}%
  \BibitemOpen
  \bibfield  {author} {\bibinfo {author} {\bibfnamefont {T.~E.}\ \bibnamefont {Li}}, \bibinfo {author} {\bibfnamefont {B.}~\bibnamefont {Cui}}, \bibinfo {author} {\bibfnamefont {J.~E.}\ \bibnamefont {Subotnik}},\ and\ \bibinfo {author} {\bibfnamefont {A.}~\bibnamefont {Nitzan}},\ }\bibfield  {title} {\bibinfo {title} {{Molecular Polaritonics: Chemical Dynamics Under Strong Light--Matter Coupling}},\ }\href {https://doi.org/10.1146/annurev-physchem-090519-042621} {\bibfield  {journal} {\bibinfo  {journal} {Annu. Rev. Phys. Chem.}\ }\textbf {\bibinfo {volume} {73}},\ \bibinfo {pages} {43} (\bibinfo {year} {2022})}\BibitemShut {NoStop}%
\bibitem [{\citenamefont {Fregoni}\ \emph {et~al.}(2022)\citenamefont {Fregoni}, \citenamefont {Garcia-Vidal},\ and\ \citenamefont {Feist}}]{Fregoni2022}%
  \BibitemOpen
  \bibfield  {author} {\bibinfo {author} {\bibfnamefont {J.}~\bibnamefont {Fregoni}}, \bibinfo {author} {\bibfnamefont {F.~J.}\ \bibnamefont {Garcia-Vidal}},\ and\ \bibinfo {author} {\bibfnamefont {J.}~\bibnamefont {Feist}},\ }\bibfield  {title} {\bibinfo {title} {{Theoretical Challenges in Polaritonic Chemistry}},\ }\href {https://doi.org/10.1021/acsphotonics.1c01749} {\bibfield  {journal} {\bibinfo  {journal} {ACS Photonics}\ }\textbf {\bibinfo {volume} {9}},\ \bibinfo {pages} {1096} (\bibinfo {year} {2022})}\BibitemShut {NoStop}%
\bibitem [{\citenamefont {Simpkins}\ \emph {et~al.}(2023)\citenamefont {Simpkins}, \citenamefont {Dunkelberger},\ and\ \citenamefont {Vurgaftman}}]{Simpkins2023}%
  \BibitemOpen
  \bibfield  {author} {\bibinfo {author} {\bibfnamefont {B.~S.}\ \bibnamefont {Simpkins}}, \bibinfo {author} {\bibfnamefont {A.~D.}\ \bibnamefont {Dunkelberger}},\ and\ \bibinfo {author} {\bibfnamefont {I.}~\bibnamefont {Vurgaftman}},\ }\bibfield  {title} {\bibinfo {title} {{Control, Modulation, and Analytical Descriptions of Vibrational Strong Coupling}},\ }\href {https://doi.org/10.1021/acs.chemrev.2c00774} {\bibfield  {journal} {\bibinfo  {journal} {Chem. Rev.}\ }\textbf {\bibinfo {volume} {123}},\ \bibinfo {pages} {5020} (\bibinfo {year} {2023})}\BibitemShut {NoStop}%
\bibitem [{\citenamefont {Mandal}\ \emph {et~al.}(2023)\citenamefont {Mandal}, \citenamefont {Taylor}, \citenamefont {Weight}, \citenamefont {Koessler}, \citenamefont {Li},\ and\ \citenamefont {Huo}}]{Mandal2023ChemRev}%
  \BibitemOpen
  \bibfield  {author} {\bibinfo {author} {\bibfnamefont {A.}~\bibnamefont {Mandal}}, \bibinfo {author} {\bibfnamefont {M.~A.}\ \bibnamefont {Taylor}}, \bibinfo {author} {\bibfnamefont {B.~M.}\ \bibnamefont {Weight}}, \bibinfo {author} {\bibfnamefont {E.~R.}\ \bibnamefont {Koessler}}, \bibinfo {author} {\bibfnamefont {X.}~\bibnamefont {Li}},\ and\ \bibinfo {author} {\bibfnamefont {P.}~\bibnamefont {Huo}},\ }\bibfield  {title} {\bibinfo {title} {{Theoretical Advances in Polariton Chemistry and Molecular Cavity Quantum Electrodynamics}},\ }\href {https://doi.org/10.1021/acs.chemrev.2c00855} {\bibfield  {journal} {\bibinfo  {journal} {Chem. Rev.}\ }\textbf {\bibinfo {volume} {123}},\ \bibinfo {pages} {9786} (\bibinfo {year} {2023})}\BibitemShut {NoStop}%
\bibitem [{\citenamefont {Ruggenthaler}\ \emph {et~al.}(2023)\citenamefont {Ruggenthaler}, \citenamefont {Sidler},\ and\ \citenamefont {Rubio}}]{Ruggenthaler2023}%
  \BibitemOpen
  \bibfield  {author} {\bibinfo {author} {\bibfnamefont {M.}~\bibnamefont {Ruggenthaler}}, \bibinfo {author} {\bibfnamefont {D.}~\bibnamefont {Sidler}},\ and\ \bibinfo {author} {\bibfnamefont {A.}~\bibnamefont {Rubio}},\ }\bibfield  {title} {\bibinfo {title} {{Understanding Polaritonic Chemistry from Ab Initio Quantum Electrodynamics}},\ }\href {https://doi.org/10.1021/acs.chemrev.2c00788} {\bibfield  {journal} {\bibinfo  {journal} {Chem. Rev.}\ }\textbf {\bibinfo {volume} {123}},\ \bibinfo {pages} {11191} (\bibinfo {year} {2023})}\BibitemShut {NoStop}%
\bibitem [{\citenamefont {Xiang}\ and\ \citenamefont {Xiong}(2024)}]{Xiang2024}%
  \BibitemOpen
  \bibfield  {author} {\bibinfo {author} {\bibfnamefont {B.}~\bibnamefont {Xiang}}\ and\ \bibinfo {author} {\bibfnamefont {W.}~\bibnamefont {Xiong}},\ }\bibfield  {title} {\bibinfo {title} {{Molecular Polaritons for Chemistry, Photonics and Quantum Technologies}},\ }\href {https://doi.org/10.1021/ACS.CHEMREV.3C00662/ASSET/IMAGES/LARGE/CR3C00662_0036.JPEG} {\bibfield  {journal} {\bibinfo  {journal} {Chem. Rev.}\ }\textbf {\bibinfo {volume} {124}},\ \bibinfo {pages} {2512} (\bibinfo {year} {2024})}\BibitemShut {NoStop}%
\bibitem [{\citenamefont {Taflove}\ and\ \citenamefont {Hagness}(2005)}]{Taflove2005}%
  \BibitemOpen
  \bibfield  {author} {\bibinfo {author} {\bibfnamefont {A.}~\bibnamefont {Taflove}}\ and\ \bibinfo {author} {\bibfnamefont {S.~C.}\ \bibnamefont {Hagness}},\ }\href@noop {} {\emph {\bibinfo {title} {{Computational Electrodynamics}}}},\ \bibinfo {edition} {3rd}\ ed.\ (\bibinfo  {publisher} {Artech House, Inc.},\ \bibinfo {address} {Norwood},\ \bibinfo {year} {2005})\BibitemShut {NoStop}%
\bibitem [{\citenamefont {Oskooi}\ \emph {et~al.}(2010)\citenamefont {Oskooi}, \citenamefont {Roundy}, \citenamefont {Ibanescu}, \citenamefont {Bermel}, \citenamefont {Joannopoulos},\ and\ \citenamefont {Johnson}}]{Oskooi2010}%
  \BibitemOpen
  \bibfield  {author} {\bibinfo {author} {\bibfnamefont {A.~F.}\ \bibnamefont {Oskooi}}, \bibinfo {author} {\bibfnamefont {D.}~\bibnamefont {Roundy}}, \bibinfo {author} {\bibfnamefont {M.}~\bibnamefont {Ibanescu}}, \bibinfo {author} {\bibfnamefont {P.}~\bibnamefont {Bermel}}, \bibinfo {author} {\bibfnamefont {J.}~\bibnamefont {Joannopoulos}},\ and\ \bibinfo {author} {\bibfnamefont {S.~G.}\ \bibnamefont {Johnson}},\ }\bibfield  {title} {\bibinfo {title} {{Meep: A Flexible Free-Software Package for Electromagnetic Simulations by the FDTD Method}},\ }\href {https://doi.org/10.1016/j.cpc.2009.11.008} {\bibfield  {journal} {\bibinfo  {journal} {Comput. Phys. Commun.}\ }\textbf {\bibinfo {volume} {181}},\ \bibinfo {pages} {687} (\bibinfo {year} {2010})}\BibitemShut {NoStop}%
\bibitem [{\citenamefont {Johansson}\ \emph {et~al.}(2012)\citenamefont {Johansson}, \citenamefont {Nation},\ and\ \citenamefont {Nori}}]{Johansson2012}%
  \BibitemOpen
  \bibfield  {author} {\bibinfo {author} {\bibfnamefont {J.}~\bibnamefont {Johansson}}, \bibinfo {author} {\bibfnamefont {P.}~\bibnamefont {Nation}},\ and\ \bibinfo {author} {\bibfnamefont {F.}~\bibnamefont {Nori}},\ }\bibfield  {title} {\bibinfo {title} {{QuTiP: An Open-Source Python Framework for the Dynamics of Open Quantum Systems}},\ }\href {https://doi.org/10.1016/j.cpc.2012.02.021} {\bibfield  {journal} {\bibinfo  {journal} {Comput. Phys. Commun.}\ }\textbf {\bibinfo {volume} {183}},\ \bibinfo {pages} {1760} (\bibinfo {year} {2012})}\BibitemShut {NoStop}%
\bibitem [{\citenamefont {{Hjorth Larsen}}\ \emph {et~al.}(2017)\citenamefont {{Hjorth Larsen}}, \citenamefont {{J{\o}rgen Mortensen}}, \citenamefont {Blomqvist}, \citenamefont {Castelli}, \citenamefont {Christensen}, \citenamefont {Du{\l}ak}, \citenamefont {Friis}, \citenamefont {Groves}, \citenamefont {Hammer}, \citenamefont {Hargus}, \citenamefont {Hermes}, \citenamefont {Jennings}, \citenamefont {{Bjerre Jensen}}, \citenamefont {Kermode}, \citenamefont {Kitchin}, \citenamefont {{Leonhard Kolsbjerg}}, \citenamefont {Kubal}, \citenamefont {Kaasbjerg}, \citenamefont {Lysgaard}, \citenamefont {{Bergmann Maronsson}}, \citenamefont {Maxson}, \citenamefont {Olsen}, \citenamefont {Pastewka}, \citenamefont {Peterson}, \citenamefont {Rostgaard}, \citenamefont {Schi{\o}tz}, \citenamefont {Sch{\"{u}}tt}, \citenamefont {Strange}, \citenamefont {Thygesen}, \citenamefont {Vegge}, \citenamefont {Vilhelmsen}, \citenamefont {Walter}, \citenamefont {Zeng},\ and\ \citenamefont {Jacobsen}}]{HjorthLarsen2017}%
  \BibitemOpen
  \bibfield  {author} {\bibinfo {author} {\bibfnamefont {A.}~\bibnamefont {{Hjorth Larsen}}}, \bibinfo {author} {\bibfnamefont {J.}~\bibnamefont {{J{\o}rgen Mortensen}}}, \bibinfo {author} {\bibfnamefont {J.}~\bibnamefont {Blomqvist}}, \bibinfo {author} {\bibfnamefont {I.~E.}\ \bibnamefont {Castelli}}, \bibinfo {author} {\bibfnamefont {R.}~\bibnamefont {Christensen}}, \bibinfo {author} {\bibfnamefont {M.}~\bibnamefont {Du{\l}ak}}, \bibinfo {author} {\bibfnamefont {J.}~\bibnamefont {Friis}}, \bibinfo {author} {\bibfnamefont {M.~N.}\ \bibnamefont {Groves}}, \bibinfo {author} {\bibfnamefont {B.}~\bibnamefont {Hammer}}, \bibinfo {author} {\bibfnamefont {C.}~\bibnamefont {Hargus}}, \bibinfo {author} {\bibfnamefont {E.~D.}\ \bibnamefont {Hermes}}, \bibinfo {author} {\bibfnamefont {P.~C.}\ \bibnamefont {Jennings}}, \bibinfo {author} {\bibfnamefont {P.}~\bibnamefont {{Bjerre Jensen}}}, \bibinfo {author} {\bibfnamefont {J.}~\bibnamefont {Kermode}}, \bibinfo {author} {\bibfnamefont {J.~R.}\ \bibnamefont {Kitchin}},
  \bibinfo {author} {\bibfnamefont {E.}~\bibnamefont {{Leonhard Kolsbjerg}}}, \bibinfo {author} {\bibfnamefont {J.}~\bibnamefont {Kubal}}, \bibinfo {author} {\bibfnamefont {K.}~\bibnamefont {Kaasbjerg}}, \bibinfo {author} {\bibfnamefont {S.}~\bibnamefont {Lysgaard}}, \bibinfo {author} {\bibfnamefont {J.}~\bibnamefont {{Bergmann Maronsson}}}, \bibinfo {author} {\bibfnamefont {T.}~\bibnamefont {Maxson}}, \bibinfo {author} {\bibfnamefont {T.}~\bibnamefont {Olsen}}, \bibinfo {author} {\bibfnamefont {L.}~\bibnamefont {Pastewka}}, \bibinfo {author} {\bibfnamefont {A.}~\bibnamefont {Peterson}}, \bibinfo {author} {\bibfnamefont {C.}~\bibnamefont {Rostgaard}}, \bibinfo {author} {\bibfnamefont {J.}~\bibnamefont {Schi{\o}tz}}, \bibinfo {author} {\bibfnamefont {O.}~\bibnamefont {Sch{\"{u}}tt}}, \bibinfo {author} {\bibfnamefont {M.}~\bibnamefont {Strange}}, \bibinfo {author} {\bibfnamefont {K.~S.}\ \bibnamefont {Thygesen}}, \bibinfo {author} {\bibfnamefont {T.}~\bibnamefont {Vegge}}, \bibinfo {author} {\bibfnamefont
  {L.}~\bibnamefont {Vilhelmsen}}, \bibinfo {author} {\bibfnamefont {M.}~\bibnamefont {Walter}}, \bibinfo {author} {\bibfnamefont {Z.}~\bibnamefont {Zeng}},\ and\ \bibinfo {author} {\bibfnamefont {K.~W.}\ \bibnamefont {Jacobsen}},\ }\bibfield  {title} {\bibinfo {title} {{The Atomic Simulation Environment -- A Python Library for Working with Atoms}},\ }\href {https://doi.org/10.1088/1361-648X/aa680e} {\bibfield  {journal} {\bibinfo  {journal} {J. Phys. Condens. Matter}\ }\textbf {\bibinfo {volume} {29}},\ \bibinfo {pages} {273002} (\bibinfo {year} {2017})}\BibitemShut {NoStop}%
\bibitem [{\citenamefont {Thompson}\ \emph {et~al.}(2022)\citenamefont {Thompson}, \citenamefont {Aktulga}, \citenamefont {Berger}, \citenamefont {Bolintineanu}, \citenamefont {Brown}, \citenamefont {Crozier}, \citenamefont {{in 't Veld}}, \citenamefont {Kohlmeyer}, \citenamefont {Moore}, \citenamefont {Nguyen}, \citenamefont {Shan}, \citenamefont {Stevens}, \citenamefont {Tranchida}, \citenamefont {Trott},\ and\ \citenamefont {Plimpton}}]{Thompson2022}%
  \BibitemOpen
  \bibfield  {author} {\bibinfo {author} {\bibfnamefont {A.~P.}\ \bibnamefont {Thompson}}, \bibinfo {author} {\bibfnamefont {H.~M.}\ \bibnamefont {Aktulga}}, \bibinfo {author} {\bibfnamefont {R.}~\bibnamefont {Berger}}, \bibinfo {author} {\bibfnamefont {D.~S.}\ \bibnamefont {Bolintineanu}}, \bibinfo {author} {\bibfnamefont {W.~M.}\ \bibnamefont {Brown}}, \bibinfo {author} {\bibfnamefont {P.~S.}\ \bibnamefont {Crozier}}, \bibinfo {author} {\bibfnamefont {P.~J.}\ \bibnamefont {{in 't Veld}}}, \bibinfo {author} {\bibfnamefont {A.}~\bibnamefont {Kohlmeyer}}, \bibinfo {author} {\bibfnamefont {S.~G.}\ \bibnamefont {Moore}}, \bibinfo {author} {\bibfnamefont {T.~D.}\ \bibnamefont {Nguyen}}, \bibinfo {author} {\bibfnamefont {R.}~\bibnamefont {Shan}}, \bibinfo {author} {\bibfnamefont {M.~J.}\ \bibnamefont {Stevens}}, \bibinfo {author} {\bibfnamefont {J.}~\bibnamefont {Tranchida}}, \bibinfo {author} {\bibfnamefont {C.}~\bibnamefont {Trott}},\ and\ \bibinfo {author} {\bibfnamefont {S.~J.}\ \bibnamefont {Plimpton}},\
  }\bibfield  {title} {\bibinfo {title} {{LAMMPS -- A Flexible Simulation Tool for Particle-based Materials Modeling at the Atomic, Neso, and Continuum Scales}},\ }\href {https://doi.org/10.1016/j.cpc.2021.108171} {\bibfield  {journal} {\bibinfo  {journal} {Comput. Phys. Commun.}\ }\textbf {\bibinfo {volume} {271}},\ \bibinfo {pages} {108171} (\bibinfo {year} {2022})}\BibitemShut {NoStop}%
\bibitem [{\citenamefont {Smith}\ \emph {et~al.}(2020)\citenamefont {Smith}, \citenamefont {Burns}, \citenamefont {Simmonett}, \citenamefont {Parrish}, \citenamefont {Schieber}, \citenamefont {Galvelis}, \citenamefont {Kraus}, \citenamefont {Kruse}, \citenamefont {{Di Remigio}}, \citenamefont {Alenaizan}, \citenamefont {James}, \citenamefont {Lehtola}, \citenamefont {Misiewicz}, \citenamefont {Scheurer}, \citenamefont {Shaw}, \citenamefont {Schriber}, \citenamefont {Xie}, \citenamefont {Glick}, \citenamefont {Sirianni}, \citenamefont {O'Brien}, \citenamefont {Waldrop}, \citenamefont {Kumar}, \citenamefont {Hohenstein}, \citenamefont {Pritchard}, \citenamefont {Brooks}, \citenamefont {Schaefer}, \citenamefont {Sokolov}, \citenamefont {Patkowski}, \citenamefont {DePrince}, \citenamefont {Bozkaya}, \citenamefont {King}, \citenamefont {Evangelista}, \citenamefont {Turney}, \citenamefont {Crawford},\ and\ \citenamefont {Sherrill}}]{Smith2020}%
  \BibitemOpen
  \bibfield  {author} {\bibinfo {author} {\bibfnamefont {D.~G.~A.}\ \bibnamefont {Smith}}, \bibinfo {author} {\bibfnamefont {L.~A.}\ \bibnamefont {Burns}}, \bibinfo {author} {\bibfnamefont {A.~C.}\ \bibnamefont {Simmonett}}, \bibinfo {author} {\bibfnamefont {R.~M.}\ \bibnamefont {Parrish}}, \bibinfo {author} {\bibfnamefont {M.~C.}\ \bibnamefont {Schieber}}, \bibinfo {author} {\bibfnamefont {R.}~\bibnamefont {Galvelis}}, \bibinfo {author} {\bibfnamefont {P.}~\bibnamefont {Kraus}}, \bibinfo {author} {\bibfnamefont {H.}~\bibnamefont {Kruse}}, \bibinfo {author} {\bibfnamefont {R.}~\bibnamefont {{Di Remigio}}}, \bibinfo {author} {\bibfnamefont {A.}~\bibnamefont {Alenaizan}}, \bibinfo {author} {\bibfnamefont {A.~M.}\ \bibnamefont {James}}, \bibinfo {author} {\bibfnamefont {S.}~\bibnamefont {Lehtola}}, \bibinfo {author} {\bibfnamefont {J.~P.}\ \bibnamefont {Misiewicz}}, \bibinfo {author} {\bibfnamefont {M.}~\bibnamefont {Scheurer}}, \bibinfo {author} {\bibfnamefont {R.~A.}\ \bibnamefont {Shaw}}, \bibinfo {author}
  {\bibfnamefont {J.~B.}\ \bibnamefont {Schriber}}, \bibinfo {author} {\bibfnamefont {Y.}~\bibnamefont {Xie}}, \bibinfo {author} {\bibfnamefont {Z.~L.}\ \bibnamefont {Glick}}, \bibinfo {author} {\bibfnamefont {D.~A.}\ \bibnamefont {Sirianni}}, \bibinfo {author} {\bibfnamefont {J.~S.}\ \bibnamefont {O'Brien}}, \bibinfo {author} {\bibfnamefont {J.~M.}\ \bibnamefont {Waldrop}}, \bibinfo {author} {\bibfnamefont {A.}~\bibnamefont {Kumar}}, \bibinfo {author} {\bibfnamefont {E.~G.}\ \bibnamefont {Hohenstein}}, \bibinfo {author} {\bibfnamefont {B.~P.}\ \bibnamefont {Pritchard}}, \bibinfo {author} {\bibfnamefont {B.~R.}\ \bibnamefont {Brooks}}, \bibinfo {author} {\bibfnamefont {H.~F.}\ \bibnamefont {Schaefer}}, \bibinfo {author} {\bibfnamefont {A.~Y.}\ \bibnamefont {Sokolov}}, \bibinfo {author} {\bibfnamefont {K.}~\bibnamefont {Patkowski}}, \bibinfo {author} {\bibfnamefont {A.~E.}\ \bibnamefont {DePrince}}, \bibinfo {author} {\bibfnamefont {U.}~\bibnamefont {Bozkaya}}, \bibinfo {author} {\bibfnamefont {R.~A.}\
  \bibnamefont {King}}, \bibinfo {author} {\bibfnamefont {F.~A.}\ \bibnamefont {Evangelista}}, \bibinfo {author} {\bibfnamefont {J.~M.}\ \bibnamefont {Turney}}, \bibinfo {author} {\bibfnamefont {T.~D.}\ \bibnamefont {Crawford}},\ and\ \bibinfo {author} {\bibfnamefont {C.~D.}\ \bibnamefont {Sherrill}},\ }\bibfield  {title} {\bibinfo {title} {{Psi4 1.4: Open-Source Software for High-Throughput Quantum Chemistry}},\ }\href {https://doi.org/10.1063/5.0006002} {\bibfield  {journal} {\bibinfo  {journal} {J. Chem. Phys.}\ }\textbf {\bibinfo {volume} {152}},\ \bibinfo {pages} {184108} (\bibinfo {year} {2020})}\BibitemShut {NoStop}%
\bibitem [{\citenamefont {Litman}\ \emph {et~al.}(2024)\citenamefont {Litman}, \citenamefont {Kapil}, \citenamefont {Feldman}, \citenamefont {Tisi}, \citenamefont {Begu{\v{s}}i{\'{c}}}, \citenamefont {Fidanyan}, \citenamefont {Fraux}, \citenamefont {Higer}, \citenamefont {Kellner}, \citenamefont {Li}, \citenamefont {P{\'{o}}s}, \citenamefont {Stocco}, \citenamefont {Trenins}, \citenamefont {Hirshberg}, \citenamefont {Rossi},\ and\ \citenamefont {Ceriotti}}]{Litman2024}%
  \BibitemOpen
  \bibfield  {author} {\bibinfo {author} {\bibfnamefont {Y.}~\bibnamefont {Litman}}, \bibinfo {author} {\bibfnamefont {V.}~\bibnamefont {Kapil}}, \bibinfo {author} {\bibfnamefont {Y.~M.}\ \bibnamefont {Feldman}}, \bibinfo {author} {\bibfnamefont {D.}~\bibnamefont {Tisi}}, \bibinfo {author} {\bibfnamefont {T.}~\bibnamefont {Begu{\v{s}}i{\'{c}}}}, \bibinfo {author} {\bibfnamefont {K.}~\bibnamefont {Fidanyan}}, \bibinfo {author} {\bibfnamefont {G.}~\bibnamefont {Fraux}}, \bibinfo {author} {\bibfnamefont {J.}~\bibnamefont {Higer}}, \bibinfo {author} {\bibfnamefont {M.}~\bibnamefont {Kellner}}, \bibinfo {author} {\bibfnamefont {T.~E.}\ \bibnamefont {Li}}, \bibinfo {author} {\bibfnamefont {E.~S.}\ \bibnamefont {P{\'{o}}s}}, \bibinfo {author} {\bibfnamefont {E.}~\bibnamefont {Stocco}}, \bibinfo {author} {\bibfnamefont {G.}~\bibnamefont {Trenins}}, \bibinfo {author} {\bibfnamefont {B.}~\bibnamefont {Hirshberg}}, \bibinfo {author} {\bibfnamefont {M.}~\bibnamefont {Rossi}},\ and\ \bibinfo {author} {\bibfnamefont
  {M.}~\bibnamefont {Ceriotti}},\ }\bibfield  {title} {\bibinfo {title} {{i-PI 3.0: A Flexible and Efficient Framework for Advanced Atomistic Simulations}},\ }\href {https://doi.org/10.1063/5.0215869} {\bibfield  {journal} {\bibinfo  {journal} {J. Chem. Phys.}\ }\textbf {\bibinfo {volume} {161}},\ \bibinfo {pages} {062504} (\bibinfo {year} {2024})}\BibitemShut {NoStop}%
\bibitem [{\citenamefont {Sukharev}\ and\ \citenamefont {Nitzan}(2017)}]{Sukharev2017}%
  \BibitemOpen
  \bibfield  {author} {\bibinfo {author} {\bibfnamefont {M.}~\bibnamefont {Sukharev}}\ and\ \bibinfo {author} {\bibfnamefont {A.}~\bibnamefont {Nitzan}},\ }\bibfield  {title} {\bibinfo {title} {{Optics of Exciton--Plasmon Nanomaterials}},\ }\href {https://doi.org/10.1088/1361-648X/aa85ef} {\bibfield  {journal} {\bibinfo  {journal} {J. Phys. Condens. Matter}\ }\textbf {\bibinfo {volume} {29}},\ \bibinfo {pages} {443003} (\bibinfo {year} {2017})}\BibitemShut {NoStop}%
\bibitem [{\citenamefont {Luk}\ \emph {et~al.}(2017)\citenamefont {Luk}, \citenamefont {Feist}, \citenamefont {Toppari},\ and\ \citenamefont {Groenhof}}]{Luk2017}%
  \BibitemOpen
  \bibfield  {author} {\bibinfo {author} {\bibfnamefont {H.~L.}\ \bibnamefont {Luk}}, \bibinfo {author} {\bibfnamefont {J.}~\bibnamefont {Feist}}, \bibinfo {author} {\bibfnamefont {J.~J.}\ \bibnamefont {Toppari}},\ and\ \bibinfo {author} {\bibfnamefont {G.}~\bibnamefont {Groenhof}},\ }\bibfield  {title} {\bibinfo {title} {{Multiscale Molecular Dynamics Simulations of Polaritonic Chemistry}},\ }\href {https://doi.org/10.1021/acs.jctc.7b00388} {\bibfield  {journal} {\bibinfo  {journal} {J. Chem. Theory Comput.}\ }\textbf {\bibinfo {volume} {13}},\ \bibinfo {pages} {4324} (\bibinfo {year} {2017})}\BibitemShut {NoStop}%
\bibitem [{\citenamefont {Li}\ \emph {et~al.}(2018{\natexlab{a}})\citenamefont {Li}, \citenamefont {Nitzan}, \citenamefont {Sukharev}, \citenamefont {Martinez}, \citenamefont {Chen},\ and\ \citenamefont {Subotnik}}]{Li2018Spontaneous}%
  \BibitemOpen
  \bibfield  {author} {\bibinfo {author} {\bibfnamefont {T.~E.}\ \bibnamefont {Li}}, \bibinfo {author} {\bibfnamefont {A.}~\bibnamefont {Nitzan}}, \bibinfo {author} {\bibfnamefont {M.}~\bibnamefont {Sukharev}}, \bibinfo {author} {\bibfnamefont {T.}~\bibnamefont {Martinez}}, \bibinfo {author} {\bibfnamefont {H.-T.}\ \bibnamefont {Chen}},\ and\ \bibinfo {author} {\bibfnamefont {J.~E.}\ \bibnamefont {Subotnik}},\ }\bibfield  {title} {\bibinfo {title} {{Mixed Quantum--Classical Electrodynamics: Understanding Spontaneous Decay and Zero--Point Energy}},\ }\href {https://doi.org/10.1103/PhysRevA.97.032105} {\bibfield  {journal} {\bibinfo  {journal} {Phys. Rev. A}\ }\textbf {\bibinfo {volume} {97}},\ \bibinfo {pages} {032105} (\bibinfo {year} {2018}{\natexlab{a}})}\BibitemShut {NoStop}%
\bibitem [{\citenamefont {Hoffmann}\ \emph {et~al.}(2018)\citenamefont {Hoffmann}, \citenamefont {Appel}, \citenamefont {Rubio},\ and\ \citenamefont {Maitra}}]{Hoffmann2018}%
  \BibitemOpen
  \bibfield  {author} {\bibinfo {author} {\bibfnamefont {N.~M.}\ \bibnamefont {Hoffmann}}, \bibinfo {author} {\bibfnamefont {H.}~\bibnamefont {Appel}}, \bibinfo {author} {\bibfnamefont {A.}~\bibnamefont {Rubio}},\ and\ \bibinfo {author} {\bibfnamefont {N.~T.}\ \bibnamefont {Maitra}},\ }\bibfield  {title} {\bibinfo {title} {{Light-matter Interactions via the Exact Factorization Approach}},\ }\href {https://doi.org/10.1140/epjb/e2018-90177-6} {\bibfield  {journal} {\bibinfo  {journal} {Eur. Phys. J. B}\ }\textbf {\bibinfo {volume} {91}},\ \bibinfo {pages} {180} (\bibinfo {year} {2018})}\BibitemShut {NoStop}%
\bibitem [{\citenamefont {Li}\ \emph {et~al.}(2020)\citenamefont {Li}, \citenamefont {Subotnik},\ and\ \citenamefont {Nitzan}}]{Li2020Water}%
  \BibitemOpen
  \bibfield  {author} {\bibinfo {author} {\bibfnamefont {T.~E.}\ \bibnamefont {Li}}, \bibinfo {author} {\bibfnamefont {J.~E.}\ \bibnamefont {Subotnik}},\ and\ \bibinfo {author} {\bibfnamefont {A.}~\bibnamefont {Nitzan}},\ }\bibfield  {title} {\bibinfo {title} {{Cavity Molecular Dynamics Simulations of Liquid Water under Vibrational Ultrastrong Coupling}},\ }\href {https://doi.org/10.1073/pnas.2009272117} {\bibfield  {journal} {\bibinfo  {journal} {Proc. Natl. Acad. Sci.}\ }\textbf {\bibinfo {volume} {117}},\ \bibinfo {pages} {18324} (\bibinfo {year} {2020})}\BibitemShut {NoStop}%
\bibitem [{\citenamefont {Sukharev}(2023)}]{Sukharev2023a}%
  \BibitemOpen
  \bibfield  {author} {\bibinfo {author} {\bibfnamefont {M.}~\bibnamefont {Sukharev}},\ }\bibfield  {title} {\bibinfo {title} {{Efficient Parallel Strategy for Molecular Plasmonics -- A Numerical Tool for Integrating Maxwell-Schr{\"{o}}dinger Equations in Three Dimensions}},\ }\href {https://doi.org/10.1016/J.JCP.2023.111920} {\bibfield  {journal} {\bibinfo  {journal} {J. Comput. Phys.}\ }\textbf {\bibinfo {volume} {477}},\ \bibinfo {pages} {111920} (\bibinfo {year} {2023})}\BibitemShut {NoStop}%
\bibitem [{\citenamefont {Xu}\ \emph {et~al.}(2023)\citenamefont {Xu}, \citenamefont {Mandal}, \citenamefont {Baxter}, \citenamefont {Cheng}, \citenamefont {Lee}, \citenamefont {Su}, \citenamefont {Liu}, \citenamefont {Reichman},\ and\ \citenamefont {Delor}}]{Xu2023Polariton}%
  \BibitemOpen
  \bibfield  {author} {\bibinfo {author} {\bibfnamefont {D.}~\bibnamefont {Xu}}, \bibinfo {author} {\bibfnamefont {A.}~\bibnamefont {Mandal}}, \bibinfo {author} {\bibfnamefont {J.~M.}\ \bibnamefont {Baxter}}, \bibinfo {author} {\bibfnamefont {S.-W.}\ \bibnamefont {Cheng}}, \bibinfo {author} {\bibfnamefont {I.}~\bibnamefont {Lee}}, \bibinfo {author} {\bibfnamefont {H.}~\bibnamefont {Su}}, \bibinfo {author} {\bibfnamefont {S.}~\bibnamefont {Liu}}, \bibinfo {author} {\bibfnamefont {D.~R.}\ \bibnamefont {Reichman}},\ and\ \bibinfo {author} {\bibfnamefont {M.}~\bibnamefont {Delor}},\ }\bibfield  {title} {\bibinfo {title} {{Ultrafast Imaging of Polariton Propagation and Interactions}},\ }\href {https://doi.org/10.1038/s41467-023-39550-x} {\bibfield  {journal} {\bibinfo  {journal} {Nat. Commun.}\ }\textbf {\bibinfo {volume} {14}},\ \bibinfo {pages} {3881} (\bibinfo {year} {2023})}\BibitemShut {NoStop}%
\bibitem [{\citenamefont {Zhou}\ \emph {et~al.}(2024)\citenamefont {Zhou}, \citenamefont {Chen}, \citenamefont {Sukharev}, \citenamefont {Subotnik},\ and\ \citenamefont {Nitzan}}]{Zhou2024}%
  \BibitemOpen
  \bibfield  {author} {\bibinfo {author} {\bibfnamefont {Z.}~\bibnamefont {Zhou}}, \bibinfo {author} {\bibfnamefont {H.~T.}\ \bibnamefont {Chen}}, \bibinfo {author} {\bibfnamefont {M.}~\bibnamefont {Sukharev}}, \bibinfo {author} {\bibfnamefont {J.~E.}\ \bibnamefont {Subotnik}},\ and\ \bibinfo {author} {\bibfnamefont {A.}~\bibnamefont {Nitzan}},\ }\bibfield  {title} {\bibinfo {title} {{Nature of Polariton Transport in a Fabry-Perot Cavity}},\ }\href {https://doi.org/10.1103/PHYSREVA.109.033717/FIGURES/6/MEDIUM} {\bibfield  {journal} {\bibinfo  {journal} {Phys. Rev. A}\ }\textbf {\bibinfo {volume} {109}},\ \bibinfo {pages} {033717} (\bibinfo {year} {2024})}\BibitemShut {NoStop}%
\bibitem [{\citenamefont {Bonaf{\'{e}}}\ \emph {et~al.}(2025)\citenamefont {Bonaf{\'{e}}}, \citenamefont {Albar}, \citenamefont {Ohlmann}, \citenamefont {Kosheleva}, \citenamefont {Bustamante}, \citenamefont {Troisi}, \citenamefont {Rubio},\ and\ \citenamefont {Appel}}]{Bonafe2025}%
  \BibitemOpen
  \bibfield  {author} {\bibinfo {author} {\bibfnamefont {F.~P.}\ \bibnamefont {Bonaf{\'{e}}}}, \bibinfo {author} {\bibfnamefont {E.~I.}\ \bibnamefont {Albar}}, \bibinfo {author} {\bibfnamefont {S.~T.}\ \bibnamefont {Ohlmann}}, \bibinfo {author} {\bibfnamefont {V.~P.}\ \bibnamefont {Kosheleva}}, \bibinfo {author} {\bibfnamefont {C.~M.}\ \bibnamefont {Bustamante}}, \bibinfo {author} {\bibfnamefont {F.}~\bibnamefont {Troisi}}, \bibinfo {author} {\bibfnamefont {A.}~\bibnamefont {Rubio}},\ and\ \bibinfo {author} {\bibfnamefont {H.}~\bibnamefont {Appel}},\ }\bibfield  {title} {\bibinfo {title} {{Full Minimal Coupling Maxwell-TDDFT: An ab initio Framework for Light-Matter Interaction beyond the Dipole Approximation}},\ }\href {https://doi.org/10.1103/PhysRevB.111.085114} {\bibfield  {journal} {\bibinfo  {journal} {Phys. Rev. B}\ }\textbf {\bibinfo {volume} {111}},\ \bibinfo {pages} {085114} (\bibinfo {year} {2025})}\BibitemShut {NoStop}%
\bibitem [{\citenamefont {Castin}\ and\ \citenamefont {Molmer}(1995)}]{Castin1995}%
  \BibitemOpen
  \bibfield  {author} {\bibinfo {author} {\bibfnamefont {Y.}~\bibnamefont {Castin}}\ and\ \bibinfo {author} {\bibfnamefont {K.}~\bibnamefont {Molmer}},\ }\bibfield  {title} {\bibinfo {title} {{Maxwell--Bloch Equations: A Unified View of Nonlinear Optics and Nonlinear Atom Optics}},\ }\href {https://doi.org/10.1103/PhysRevA.51.R3426} {\bibfield  {journal} {\bibinfo  {journal} {Phys. Rev. A}\ }\textbf {\bibinfo {volume} {51}},\ \bibinfo {pages} {R3426} (\bibinfo {year} {1995})}\BibitemShut {NoStop}%
\bibitem [{\citenamefont {Lopata}\ and\ \citenamefont {Neuhauser}(2009{\natexlab{a}})}]{Lopata2009-1}%
  \BibitemOpen
  \bibfield  {author} {\bibinfo {author} {\bibfnamefont {K.}~\bibnamefont {Lopata}}\ and\ \bibinfo {author} {\bibfnamefont {D.}~\bibnamefont {Neuhauser}},\ }\bibfield  {title} {\bibinfo {title} {{Multiscale Maxwell--Schr{\"{o}}dinger Modeling: A Split Field Finite--Difference Time--Domain Approach to Molecular Nanopolaritonics}},\ }\href {https://doi.org/10.1063/1.3082245} {\bibfield  {journal} {\bibinfo  {journal} {J. Chem. Phys.}\ }\textbf {\bibinfo {volume} {130}},\ \bibinfo {pages} {104707} (\bibinfo {year} {2009}{\natexlab{a}})}\BibitemShut {NoStop}%
\bibitem [{\citenamefont {Lopata}\ and\ \citenamefont {Neuhauser}(2009{\natexlab{b}})}]{Lopata2009-2}%
  \BibitemOpen
  \bibfield  {author} {\bibinfo {author} {\bibfnamefont {K.}~\bibnamefont {Lopata}}\ and\ \bibinfo {author} {\bibfnamefont {D.}~\bibnamefont {Neuhauser}},\ }\bibfield  {title} {\bibinfo {title} {{Nonlinear Nanopolaritonics: Finite--Difference Time--Domain Maxwell--Schr{\"{o}}dinger Simulation of Molecule-Assisted Plasmon Transfer}},\ }\href {https://doi.org/10.1063/1.3167407} {\bibfield  {journal} {\bibinfo  {journal} {J. Chem. Phys.}\ }\textbf {\bibinfo {volume} {131}},\ \bibinfo {pages} {014701} (\bibinfo {year} {2009}{\natexlab{b}})}\BibitemShut {NoStop}%
\bibitem [{\citenamefont {Chen}\ \emph {et~al.}(2010)\citenamefont {Chen}, \citenamefont {McMahon}, \citenamefont {Ratner},\ and\ \citenamefont {Schatz}}]{Chen2010}%
  \BibitemOpen
  \bibfield  {author} {\bibinfo {author} {\bibfnamefont {H.}~\bibnamefont {Chen}}, \bibinfo {author} {\bibfnamefont {J.~M.}\ \bibnamefont {McMahon}}, \bibinfo {author} {\bibfnamefont {M.~A.}\ \bibnamefont {Ratner}},\ and\ \bibinfo {author} {\bibfnamefont {G.~C.}\ \bibnamefont {Schatz}},\ }\bibfield  {title} {\bibinfo {title} {{Classical Electrodynamics Coupled to Quantum Mechanics for Calculation of Molecular Optical Properties: a RT-TDDFT/FDTD Approach}},\ }\href {https://doi.org/10.1021/jp1043392} {\bibfield  {journal} {\bibinfo  {journal} {J. Phys. Chem. C}\ }\textbf {\bibinfo {volume} {114}},\ \bibinfo {pages} {14384} (\bibinfo {year} {2010})}\BibitemShut {NoStop}%
\bibitem [{\citenamefont {Sidler}\ \emph {et~al.}(2025)\citenamefont {Sidler}, \citenamefont {Bustamante}, \citenamefont {Bonaf{\'{e}}}, \citenamefont {Ruggenthaler}, \citenamefont {Sukharev},\ and\ \citenamefont {Rubio}}]{Sidler2025}%
  \BibitemOpen
  \bibfield  {author} {\bibinfo {author} {\bibfnamefont {D.}~\bibnamefont {Sidler}}, \bibinfo {author} {\bibfnamefont {C.~M.}\ \bibnamefont {Bustamante}}, \bibinfo {author} {\bibfnamefont {F.~P.}\ \bibnamefont {Bonaf{\'{e}}}}, \bibinfo {author} {\bibfnamefont {M.}~\bibnamefont {Ruggenthaler}}, \bibinfo {author} {\bibfnamefont {M.}~\bibnamefont {Sukharev}},\ and\ \bibinfo {author} {\bibfnamefont {A.}~\bibnamefont {Rubio}},\ }\bibfield  {title} {\bibinfo {title} {{Density-Functional Tight Binding Meets Maxwell: Unraveling the Mysteries of (Strong) Light-Matter Coupling Efficiently}},\ }\href {https://doi.org/10.1515/nanoph-2025-0453} {\bibfield  {journal} {\bibinfo  {journal} {Nanophotonics}\ ,\ \bibinfo {pages} {1}} (\bibinfo {year} {2025})}\BibitemShut {NoStop}%
\bibitem [{\citenamefont {Yamada}(2020)}]{Yamada2020}%
  \BibitemOpen
  \bibfield  {author} {\bibinfo {author} {\bibfnamefont {A.}~\bibnamefont {Yamada}},\ }\bibfield  {title} {\bibinfo {title} {{Multiscale Coupled Maxwell's Equations and Polarizable Molecular Dynamics Simulation based on Charge Response Kernel Model}},\ }\href {https://doi.org/10.1063/1.5143742/1063053} {\bibfield  {journal} {\bibinfo  {journal} {J. Chem. Phys.}\ }\textbf {\bibinfo {volume} {152}},\ \bibinfo {pages} {94110} (\bibinfo {year} {2020})}\BibitemShut {NoStop}%
\bibitem [{\citenamefont {Li}(2024)}]{Li2024CavMD}%
  \BibitemOpen
  \bibfield  {author} {\bibinfo {author} {\bibfnamefont {T.~E.}\ \bibnamefont {Li}},\ }\bibfield  {title} {\bibinfo {title} {{Mesoscale Molecular Simulations of Fabry-P{\'{e}}rot Vibrational Strong Coupling}},\ }\href {https://doi.org/10.1021/acs.jctc.4c00349} {\bibfield  {journal} {\bibinfo  {journal} {J. Chem. Theory Comput.}\ }\textbf {\bibinfo {volume} {20}},\ \bibinfo {pages} {7016} (\bibinfo {year} {2024})}\BibitemShut {NoStop}%
\bibitem [{\citenamefont {Sokolovskii}\ \emph {et~al.}(2023)\citenamefont {Sokolovskii}, \citenamefont {Tichauer}, \citenamefont {Morozov}, \citenamefont {Feist},\ and\ \citenamefont {Groenhof}}]{Sokolovskii2022tmp}%
  \BibitemOpen
  \bibfield  {author} {\bibinfo {author} {\bibfnamefont {I.}~\bibnamefont {Sokolovskii}}, \bibinfo {author} {\bibfnamefont {R.~H.}\ \bibnamefont {Tichauer}}, \bibinfo {author} {\bibfnamefont {D.}~\bibnamefont {Morozov}}, \bibinfo {author} {\bibfnamefont {J.}~\bibnamefont {Feist}},\ and\ \bibinfo {author} {\bibfnamefont {G.}~\bibnamefont {Groenhof}},\ }\bibfield  {title} {\bibinfo {title} {{Multi-scale Molecular Dynamics Simulations of Enhanced Energy Transfer in Organic Molecules under Strong Coupling}},\ }\href {https://doi.org/10.1038/s41467-023-42067-y} {\bibfield  {journal} {\bibinfo  {journal} {Nat. Commun.}\ }\textbf {\bibinfo {volume} {14}},\ \bibinfo {pages} {6613} (\bibinfo {year} {2023})}\BibitemShut {NoStop}%
\bibitem [{\citenamefont {Nguyen}\ and\ \citenamefont {Ribeiro}(2025)}]{Nguyen2025}%
  \BibitemOpen
  \bibfield  {author} {\bibinfo {author} {\bibfnamefont {T.~H.}\ \bibnamefont {Nguyen}}\ and\ \bibinfo {author} {\bibfnamefont {R.~F.}\ \bibnamefont {Ribeiro}},\ }\bibfield  {title} {\bibinfo {title} {{Collision-Induced Spectroscopy and Radiative Association in Microcavities}},\ }\href {https://doi.org/10.1063/5.0271753/3353198} {\bibfield  {journal} {\bibinfo  {journal} {J. Chem. Phys.}\ }\textbf {\bibinfo {volume} {163}},\ \bibinfo {pages} {34114} (\bibinfo {year} {2025})}\BibitemShut {NoStop}%
\bibitem [{\citenamefont {Bruner}\ \emph {et~al.}(2016)\citenamefont {Bruner}, \citenamefont {Lamaster},\ and\ \citenamefont {Lopata}}]{Bruner2016}%
  \BibitemOpen
  \bibfield  {author} {\bibinfo {author} {\bibfnamefont {A.}~\bibnamefont {Bruner}}, \bibinfo {author} {\bibfnamefont {D.}~\bibnamefont {Lamaster}},\ and\ \bibinfo {author} {\bibfnamefont {K.}~\bibnamefont {Lopata}},\ }\bibfield  {title} {\bibinfo {title} {{Accelerated Broadband Spectra Using Transition Dipole Decomposition and Pad{\'{e}} Approximants}},\ }\href {https://doi.org/10.1021/ACS.JCTC.6B00511/SUPPL_FILE/CT6B00511_SI_001.PDF} {\bibfield  {journal} {\bibinfo  {journal} {J. Chem. Theory Comput.}\ }\textbf {\bibinfo {volume} {12}},\ \bibinfo {pages} {3741} (\bibinfo {year} {2016})}\BibitemShut {NoStop}%
\bibitem [{\citenamefont {Dar}\ \emph {et~al.}(2024)\citenamefont {Dar}, \citenamefont {Baranova},\ and\ \citenamefont {Maitra}}]{Dar2024}%
  \BibitemOpen
  \bibfield  {author} {\bibinfo {author} {\bibfnamefont {D.~B.}\ \bibnamefont {Dar}}, \bibinfo {author} {\bibfnamefont {A.}~\bibnamefont {Baranova}},\ and\ \bibinfo {author} {\bibfnamefont {N.~T.}\ \bibnamefont {Maitra}},\ }\bibfield  {title} {\bibinfo {title} {{Reformulation of Time-Dependent Density Functional Theory for Nonperturbative Dynamics: The Rabi Oscillation Problem Resolved}},\ }\href {https://doi.org/10.1103/PhysRevLett.133.096401} {\bibfield  {journal} {\bibinfo  {journal} {Phys. Rev. Lett.}\ }\textbf {\bibinfo {volume} {133}},\ \bibinfo {pages} {096401} (\bibinfo {year} {2024})}\BibitemShut {NoStop}%
\bibitem [{\citenamefont {Li}\ \emph {et~al.}(2005)\citenamefont {Li}, \citenamefont {Tully}, \citenamefont {Schlegel},\ and\ \citenamefont {Frisch}}]{Li2005Eh}%
  \BibitemOpen
  \bibfield  {author} {\bibinfo {author} {\bibfnamefont {X.}~\bibnamefont {Li}}, \bibinfo {author} {\bibfnamefont {J.~C.}\ \bibnamefont {Tully}}, \bibinfo {author} {\bibfnamefont {H.~B.}\ \bibnamefont {Schlegel}},\ and\ \bibinfo {author} {\bibfnamefont {M.~J.}\ \bibnamefont {Frisch}},\ }\bibfield  {title} {\bibinfo {title} {{Ab initio Ehrenfest dynamics}},\ }\href {https://doi.org/10.1063/1.2008258} {\bibfield  {journal} {\bibinfo  {journal} {J. Chem. Phys.}\ }\textbf {\bibinfo {volume} {123}},\ \bibinfo {pages} {084106} (\bibinfo {year} {2005})}\BibitemShut {NoStop}%
\bibitem [{\citenamefont {Goings}\ \emph {et~al.}(2018)\citenamefont {Goings}, \citenamefont {Lestrange},\ and\ \citenamefont {Li}}]{Goings2018}%
  \BibitemOpen
  \bibfield  {author} {\bibinfo {author} {\bibfnamefont {J.~J.}\ \bibnamefont {Goings}}, \bibinfo {author} {\bibfnamefont {P.~J.}\ \bibnamefont {Lestrange}},\ and\ \bibinfo {author} {\bibfnamefont {X.}~\bibnamefont {Li}},\ }\bibfield  {title} {\bibinfo {title} {{Real-Time Time-Dependent Electronic Structure Theory}},\ }\href {https://doi.org/10.1002/WCMS.1341} {\bibfield  {journal} {\bibinfo  {journal} {Wiley Interdiscip. Rev.: Comput. Mol. Sci.}\ }\textbf {\bibinfo {volume} {8}},\ \bibinfo {pages} {e1341} (\bibinfo {year} {2018})}\BibitemShut {NoStop}%
\bibitem [{Max()}]{MaxwellLinkDocument}%
  \BibitemOpen
  \href@noop {} {\bibinfo {title} {{MaxwellLink Documentation Website}}},\ \bibinfo {howpublished} {\url{https://taoeli.github.io/MaxwellLink/}},\ \bibinfo {note} {accessed: 2025-12-05}\BibitemShut {NoStop}%
\bibitem [{\citenamefont {Dicke}(1954)}]{Dicke1954}%
  \BibitemOpen
  \bibfield  {author} {\bibinfo {author} {\bibfnamefont {R.~H.}\ \bibnamefont {Dicke}},\ }\bibfield  {title} {\bibinfo {title} {{Coherence in Spontaneous Radiation Processes}},\ }\href {https://doi.org/10.1103/PhysRev.93.99} {\bibfield  {journal} {\bibinfo  {journal} {Phys. Rev.}\ }\textbf {\bibinfo {volume} {93}},\ \bibinfo {pages} {99} (\bibinfo {year} {1954})}\BibitemShut {NoStop}%
\bibitem [{\citenamefont {Gross}\ and\ \citenamefont {Haroche}(1982)}]{Gross1982}%
  \BibitemOpen
  \bibfield  {author} {\bibinfo {author} {\bibfnamefont {M.}~\bibnamefont {Gross}}\ and\ \bibinfo {author} {\bibfnamefont {S.}~\bibnamefont {Haroche}},\ }\bibfield  {title} {\bibinfo {title} {{Superradiance: An Essay on the Theory of Collective Spontaneous Emission}},\ }\href {https://doi.org/10.1016/0370-1573(82)90102-8} {\bibfield  {journal} {\bibinfo  {journal} {Phys. Rep.}\ }\textbf {\bibinfo {volume} {93}},\ \bibinfo {pages} {301} (\bibinfo {year} {1982})}\BibitemShut {NoStop}%
\bibitem [{\citenamefont {Griffiths}(1999)}]{Griffiths1999}%
  \BibitemOpen
  \bibfield  {author} {\bibinfo {author} {\bibfnamefont {D.~J.}\ \bibnamefont {Griffiths}},\ }\href@noop {} {\emph {\bibinfo {title} {{Introduction to Electrodynamics}}}},\ \bibinfo {edition} {3rd}\ ed.\ (\bibinfo  {publisher} {Prentice-Hall, Inc.},\ \bibinfo {address} {New Jersey},\ \bibinfo {year} {1999})\BibitemShut {NoStop}%
\bibitem [{\citenamefont {Yee}(1966)}]{Yee1966}%
  \BibitemOpen
  \bibfield  {author} {\bibinfo {author} {\bibfnamefont {K.}~\bibnamefont {Yee}},\ }\bibfield  {title} {\bibinfo {title} {{Numerical Solution of Initial Boundary Value Problems Involving Maxwell's Equations in Isotropic Media}},\ }\href {https://doi.org/10.1109/TAP.1966.1138693} {\bibfield  {journal} {\bibinfo  {journal} {Antennas Propagation, IEEE Trans.}\ }\textbf {\bibinfo {volume} {14}},\ \bibinfo {pages} {302} (\bibinfo {year} {1966})}\BibitemShut {NoStop}%
\bibitem [{\citenamefont {Cohen-Tannoudji}\ \emph {et~al.}(1997)\citenamefont {Cohen-Tannoudji}, \citenamefont {Dupont-Roc},\ and\ \citenamefont {Grynberg}}]{Cohen-Tannoudji1997}%
  \BibitemOpen
  \bibfield  {author} {\bibinfo {author} {\bibfnamefont {C.}~\bibnamefont {Cohen-Tannoudji}}, \bibinfo {author} {\bibfnamefont {J.}~\bibnamefont {Dupont-Roc}},\ and\ \bibinfo {author} {\bibfnamefont {G.}~\bibnamefont {Grynberg}},\ }\href@noop {} {\emph {\bibinfo {title} {{Photons and Atoms: Introduction to Quantum Electrodynamics}}}}\ (\bibinfo  {publisher} {Wiley},\ \bibinfo {address} {New York},\ \bibinfo {year} {1997})\ pp.\ \bibinfo {pages} {280--295}\BibitemShut {NoStop}%
\bibitem [{\citenamefont {Novotny}\ and\ \citenamefont {Hecht}(2006)}]{Novotny2006}%
  \BibitemOpen
  \bibfield  {author} {\bibinfo {author} {\bibfnamefont {L.}~\bibnamefont {Novotny}}\ and\ \bibinfo {author} {\bibfnamefont {B.}~\bibnamefont {Hecht}},\ }\href@noop {} {\emph {\bibinfo {title} {{Principles of Nano-Optics}}}}\ (\bibinfo  {publisher} {Cambridge University Press},\ \bibinfo {address} {Cambridge},\ \bibinfo {year} {2006})\BibitemShut {NoStop}%
\bibitem [{\citenamefont {Ding}\ \emph {et~al.}(2017)\citenamefont {Ding}, \citenamefont {Hsu},\ and\ \citenamefont {Schatz}}]{Ding2017}%
  \BibitemOpen
  \bibfield  {author} {\bibinfo {author} {\bibfnamefont {W.}~\bibnamefont {Ding}}, \bibinfo {author} {\bibfnamefont {L.-Y.}\ \bibnamefont {Hsu}},\ and\ \bibinfo {author} {\bibfnamefont {G.~C.}\ \bibnamefont {Schatz}},\ }\bibfield  {title} {\bibinfo {title} {{Plasmon-Coupled Resonance Energy Transfer: A Real-Time Electrodynamics Approach}},\ }\href {https://doi.org/10.1063/1.4975815} {\bibfield  {journal} {\bibinfo  {journal} {J. Chem. Phys.}\ }\textbf {\bibinfo {volume} {146}},\ \bibinfo {pages} {064109} (\bibinfo {year} {2017})}\BibitemShut {NoStop}%
\bibitem [{\citenamefont {Smith}\ \emph {et~al.}(1969)\citenamefont {Smith}, \citenamefont {Vidal},\ and\ \citenamefont {Cooper}}]{Smith1969}%
  \BibitemOpen
  \bibfield  {author} {\bibinfo {author} {\bibfnamefont {E.~W.}\ \bibnamefont {Smith}}, \bibinfo {author} {\bibfnamefont {C.}~\bibnamefont {Vidal}},\ and\ \bibinfo {author} {\bibfnamefont {J.}~\bibnamefont {Cooper}},\ }\bibfield  {title} {\bibinfo {title} {{Classical Path Methods in Line Broadening. I. The Classical Path Approximation}},\ }\href {https://doi.org/10.6028/jres.073A.030} {\bibfield  {journal} {\bibinfo  {journal} {J. Res. Natl. Bur. Stand. A. Phys. Chem.}\ }\textbf {\bibinfo {volume} {73A}},\ \bibinfo {pages} {389} (\bibinfo {year} {1969})}\BibitemShut {NoStop}%
\bibitem [{\citenamefont {Zhao}\ \emph {et~al.}(2020)\citenamefont {Zhao}, \citenamefont {Wildman}, \citenamefont {Tao}, \citenamefont {Schneider}, \citenamefont {Hammes-Schiffer},\ and\ \citenamefont {Li}}]{Zhao2020JCP}%
  \BibitemOpen
  \bibfield  {author} {\bibinfo {author} {\bibfnamefont {L.}~\bibnamefont {Zhao}}, \bibinfo {author} {\bibfnamefont {A.}~\bibnamefont {Wildman}}, \bibinfo {author} {\bibfnamefont {Z.}~\bibnamefont {Tao}}, \bibinfo {author} {\bibfnamefont {P.}~\bibnamefont {Schneider}}, \bibinfo {author} {\bibfnamefont {S.}~\bibnamefont {Hammes-Schiffer}},\ and\ \bibinfo {author} {\bibfnamefont {X.}~\bibnamefont {Li}},\ }\bibfield  {title} {\bibinfo {title} {{Nuclear-Electronic Orbital Ehrenfest Dynamics}},\ }\href {https://doi.org/10.1063/5.0031019} {\bibfield  {journal} {\bibinfo  {journal} {J. Chem. Phys.}\ }\textbf {\bibinfo {volume} {153}},\ \bibinfo {pages} {224111} (\bibinfo {year} {2020})}\BibitemShut {NoStop}%
\bibitem [{\citenamefont {Li}\ and\ \citenamefont {Hammes-Schiffer}(2023)}]{Li2023eBO}%
  \BibitemOpen
  \bibfield  {author} {\bibinfo {author} {\bibfnamefont {T.~E.}\ \bibnamefont {Li}}\ and\ \bibinfo {author} {\bibfnamefont {S.}~\bibnamefont {Hammes-Schiffer}},\ }\bibfield  {title} {\bibinfo {title} {{Electronic Born-Oppenheimer Approximation in Nuclear-Electronic Orbital Dynamics}},\ }\href {https://doi.org/10.1063/5.0142007/2881560} {\bibfield  {journal} {\bibinfo  {journal} {J. Chem. Phys.}\ }\textbf {\bibinfo {volume} {158}},\ \bibinfo {pages} {114118} (\bibinfo {year} {2023})}\BibitemShut {NoStop}%
\bibitem [{\citenamefont {{Bocanegra Vargas}}\ and\ \citenamefont {Li}(2025)}]{Vargas2024}%
  \BibitemOpen
  \bibfield  {author} {\bibinfo {author} {\bibfnamefont {A.~F.}\ \bibnamefont {{Bocanegra Vargas}}}\ and\ \bibinfo {author} {\bibfnamefont {T.~E.}\ \bibnamefont {Li}},\ }\bibfield  {title} {\bibinfo {title} {{Polariton-Induced Purcell effects via a Reduced Semiclassical Electrodynamics Approach}},\ }\href {https://doi.org/10.1063/5.0251767/3340319} {\bibfield  {journal} {\bibinfo  {journal} {J. Chem. Phys.}\ }\textbf {\bibinfo {volume} {162}},\ \bibinfo {pages} {124101} (\bibinfo {year} {2025})}\BibitemShut {NoStop}%
\bibitem [{\citenamefont {Jaynes}\ and\ \citenamefont {Cummings}(1963)}]{Jaynes1963}%
  \BibitemOpen
  \bibfield  {author} {\bibinfo {author} {\bibfnamefont {E.}~\bibnamefont {Jaynes}}\ and\ \bibinfo {author} {\bibfnamefont {F.}~\bibnamefont {Cummings}},\ }\bibfield  {title} {\bibinfo {title} {{Comparison of Quantum and Semiclassical Radiation Theories with Application to the Beam Maser}},\ }\href {https://doi.org/10.1109/PROC.1963.1664} {\bibfield  {journal} {\bibinfo  {journal} {Proc. IEEE}\ }\textbf {\bibinfo {volume} {51}},\ \bibinfo {pages} {89} (\bibinfo {year} {1963})}\BibitemShut {NoStop}%
\bibitem [{\citenamefont {Chen}\ \emph {et~al.}(2019)\citenamefont {Chen}, \citenamefont {Li}, \citenamefont {Sukharev}, \citenamefont {Nitzan},\ and\ \citenamefont {Subotnik}}]{Chen2018Spontaneous}%
  \BibitemOpen
  \bibfield  {author} {\bibinfo {author} {\bibfnamefont {H.-T.}\ \bibnamefont {Chen}}, \bibinfo {author} {\bibfnamefont {T.~E.}\ \bibnamefont {Li}}, \bibinfo {author} {\bibfnamefont {M.}~\bibnamefont {Sukharev}}, \bibinfo {author} {\bibfnamefont {A.}~\bibnamefont {Nitzan}},\ and\ \bibinfo {author} {\bibfnamefont {J.~E.}\ \bibnamefont {Subotnik}},\ }\bibfield  {title} {\bibinfo {title} {{Ehrenfest+R dynamics. I. A Mixed Quantum-Classical Electrodynamics Simulation of Spontaneous Emission}},\ }\href {https://doi.org/10.1063/1.5057365} {\bibfield  {journal} {\bibinfo  {journal} {J. Chem. Phys.}\ }\textbf {\bibinfo {volume} {150}},\ \bibinfo {pages} {044102} (\bibinfo {year} {2019})}\BibitemShut {NoStop}%
\bibitem [{\citenamefont {Boerner}\ \emph {et~al.}(2023)\citenamefont {Boerner}, \citenamefont {Deems}, \citenamefont {Furlani}, \citenamefont {Knuth},\ and\ \citenamefont {Towns}}]{Boerner2023}%
  \BibitemOpen
  \bibfield  {author} {\bibinfo {author} {\bibfnamefont {T.~J.}\ \bibnamefont {Boerner}}, \bibinfo {author} {\bibfnamefont {S.}~\bibnamefont {Deems}}, \bibinfo {author} {\bibfnamefont {T.~R.}\ \bibnamefont {Furlani}}, \bibinfo {author} {\bibfnamefont {S.~L.}\ \bibnamefont {Knuth}},\ and\ \bibinfo {author} {\bibfnamefont {J.}~\bibnamefont {Towns}},\ }\bibfield  {title} {\bibinfo {title} {{ACCESS: Advancing Innovation: NSF's Advanced Cyberinfrastructure Coordination Ecosystem: Services \& Support}},\ }in\ \href {https://doi.org/10.1145/3569951.3597559} {\emph {\bibinfo {booktitle} {Practice and Experience in Advanced Research Computing}}}\ (\bibinfo  {publisher} {ACM},\ \bibinfo {address} {New York, NY, USA},\ \bibinfo {year} {2023})\ pp.\ \bibinfo {pages} {173--176}\BibitemShut {NoStop}%
\bibitem [{\citenamefont {Li}\ \emph {et~al.}(2018{\natexlab{b}})\citenamefont {Li}, \citenamefont {Chen}, \citenamefont {Nitzan}, \citenamefont {Sukharev},\ and\ \citenamefont {Subotnik}}]{Li2018Tradeoff}%
  \BibitemOpen
  \bibfield  {author} {\bibinfo {author} {\bibfnamefont {T.~E.}\ \bibnamefont {Li}}, \bibinfo {author} {\bibfnamefont {H.-T.}\ \bibnamefont {Chen}}, \bibinfo {author} {\bibfnamefont {A.}~\bibnamefont {Nitzan}}, \bibinfo {author} {\bibfnamefont {M.}~\bibnamefont {Sukharev}},\ and\ \bibinfo {author} {\bibfnamefont {J.~E.}\ \bibnamefont {Subotnik}},\ }\bibfield  {title} {\bibinfo {title} {{A Necessary Trade-off for Semiclassical Electrodynamics: Accurate Short-Range Coulomb Interactions versus the Enforcement of Causality?}},\ }\href {https://doi.org/10.1021/acs.jpclett.8b02309} {\bibfield  {journal} {\bibinfo  {journal} {J. Phys. Chem. Lett.}\ ,\ \bibinfo {pages} {5955}} (\bibinfo {year} {2018}{\natexlab{b}})}\BibitemShut {NoStop}%
\bibitem [{\citenamefont {Salam}(2018)}]{Salam2018}%
  \BibitemOpen
  \bibfield  {author} {\bibinfo {author} {\bibfnamefont {A.}~\bibnamefont {Salam}},\ }\bibfield  {title} {\bibinfo {title} {{The Unified Theory of Resonance Energy Transfer According to Molecular Quantum Electrodynamics}},\ }\href {https://doi.org/10.3390/atoms6040056} {\bibfield  {journal} {\bibinfo  {journal} {Atoms}\ }\textbf {\bibinfo {volume} {6}},\ \bibinfo {pages} {56} (\bibinfo {year} {2018})}\BibitemShut {NoStop}%
\bibitem [{\citenamefont {Lee}\ \emph {et~al.}(1988)\citenamefont {Lee}, \citenamefont {Yang},\ and\ \citenamefont {Parr}}]{Lee1988}%
  \BibitemOpen
  \bibfield  {author} {\bibinfo {author} {\bibfnamefont {C.}~\bibnamefont {Lee}}, \bibinfo {author} {\bibfnamefont {W.}~\bibnamefont {Yang}},\ and\ \bibinfo {author} {\bibfnamefont {R.~G.}\ \bibnamefont {Parr}},\ }\bibfield  {title} {\bibinfo {title} {{Development of the Colle--Salvetti Correlation-Energy Formula into a Functional of the Electron Density}},\ }\href {https://doi.org/10.1103/PhysRevB.37.785} {\bibfield  {journal} {\bibinfo  {journal} {Phys. Rev. B}\ }\textbf {\bibinfo {volume} {37}},\ \bibinfo {pages} {785} (\bibinfo {year} {1988})}\BibitemShut {NoStop}%
\bibitem [{\citenamefont {Becke}(1988)}]{Becke1988}%
  \BibitemOpen
  \bibfield  {author} {\bibinfo {author} {\bibfnamefont {A.~D.}\ \bibnamefont {Becke}},\ }\bibfield  {title} {\bibinfo {title} {{Density-Functional Exchange-Energy Approximation with Correct Asymptotic Behavior}},\ }\href {https://doi.org/10.1103/PhysRevA.38.3098} {\bibfield  {journal} {\bibinfo  {journal} {Phys. Rev. A}\ }\textbf {\bibinfo {volume} {38}},\ \bibinfo {pages} {3098} (\bibinfo {year} {1988})}\BibitemShut {NoStop}%
\bibitem [{\citenamefont {Becke}(1998)}]{Becke1998}%
  \BibitemOpen
  \bibfield  {author} {\bibinfo {author} {\bibfnamefont {A.~D.}\ \bibnamefont {Becke}},\ }\bibfield  {title} {\bibinfo {title} {{A New Inhomogeneity Parameter in Density-Functional Theory}},\ }\href {https://doi.org/10.1063/1.476722} {\bibfield  {journal} {\bibinfo  {journal} {J. Chem. Phys.}\ }\textbf {\bibinfo {volume} {109}},\ \bibinfo {pages} {2092} (\bibinfo {year} {1998})}\BibitemShut {NoStop}%
\bibitem [{\citenamefont {Dunning}(1989)}]{Dunning1989}%
  \BibitemOpen
  \bibfield  {author} {\bibinfo {author} {\bibfnamefont {T.~H.}\ \bibnamefont {Dunning}},\ }\bibfield  {title} {\bibinfo {title} {{Gaussian Basis Sets for Use in Correlated Molecular Calculations. I. The Atoms Boron Through Neon and Hydrogen}},\ }\href {https://doi.org/10.1063/1.456153} {\bibfield  {journal} {\bibinfo  {journal} {J. Chem. Phys.}\ }\textbf {\bibinfo {volume} {90}},\ \bibinfo {pages} {1007} (\bibinfo {year} {1989})}\BibitemShut {NoStop}%
\bibitem [{\citenamefont {Li}\ and\ \citenamefont {Tong}(1986)}]{Li1986TDDFTMulti}%
  \BibitemOpen
  \bibfield  {author} {\bibinfo {author} {\bibfnamefont {T.-C.}\ \bibnamefont {Li}}\ and\ \bibinfo {author} {\bibfnamefont {P.-Q.}\ \bibnamefont {Tong}},\ }\bibfield  {title} {\bibinfo {title} {{Time-Dependent Density-Functional Theory for Multicomponent Systems}},\ }\href {https://doi.org/10.1103/physreva.34.529} {\bibfield  {journal} {\bibinfo  {journal} {Phys. Rev. A}\ }\textbf {\bibinfo {volume} {34}},\ \bibinfo {pages} {529} (\bibinfo {year} {1986})}\BibitemShut {NoStop}%
\bibitem [{\citenamefont {van Leeuwen}\ and\ \citenamefont {Gross}(2006)}]{Marques2006TDDFT}%
  \BibitemOpen
  \bibfield  {author} {\bibinfo {author} {\bibfnamefont {R.}~\bibnamefont {van Leeuwen}}\ and\ \bibinfo {author} {\bibfnamefont {E.~K.~U.}\ \bibnamefont {Gross}},\ }\bibfield  {title} {\bibinfo {title} {Multicomponent density-functional theory},\ }in\ \href {https://doi.org/10.1007/b11767107} {\emph {\bibinfo {booktitle} {Time-Dependent Density Functional Theory}}},\ \bibinfo {editor} {edited by\ \bibinfo {editor} {\bibfnamefont {M.~A.}\ \bibnamefont {Marques}}, \bibinfo {editor} {\bibfnamefont {C.~A.}\ \bibnamefont {Ullrich}}, \bibinfo {editor} {\bibfnamefont {F.}~\bibnamefont {Nogueira}}, \bibinfo {editor} {\bibfnamefont {A.}~\bibnamefont {Rubio}}, \bibinfo {editor} {\bibfnamefont {K.}~\bibnamefont {Burke}},\ and\ \bibinfo {editor} {\bibfnamefont {E.~K.~U.}\ \bibnamefont {Gross}}}\ (\bibinfo  {publisher} {Springer Berlin Heidelberg},\ \bibinfo {year} {2006})\ pp.\ \bibinfo {pages} {93--106}\BibitemShut {NoStop}%
\bibitem [{\citenamefont {Butriy}\ \emph {et~al.}(2007)\citenamefont {Butriy}, \citenamefont {Ebadi}, \citenamefont {de~Boeij}, \citenamefont {van Leeuwen},\ and\ \citenamefont {Gross}}]{Butriy2007}%
  \BibitemOpen
  \bibfield  {author} {\bibinfo {author} {\bibfnamefont {O.}~\bibnamefont {Butriy}}, \bibinfo {author} {\bibfnamefont {H.}~\bibnamefont {Ebadi}}, \bibinfo {author} {\bibfnamefont {P.~L.}\ \bibnamefont {de~Boeij}}, \bibinfo {author} {\bibfnamefont {R.}~\bibnamefont {van Leeuwen}},\ and\ \bibinfo {author} {\bibfnamefont {E.~K.~U.}\ \bibnamefont {Gross}},\ }\bibfield  {title} {\bibinfo {title} {{Multicomponent Density-Functional Theory for Time-Dependent Systems}},\ }\href {https://doi.org/10.1103/PhysRevA.76.052514} {\bibfield  {journal} {\bibinfo  {journal} {Phys. Rev. A}\ }\textbf {\bibinfo {volume} {76}},\ \bibinfo {pages} {052514} (\bibinfo {year} {2007})}\BibitemShut {NoStop}%
\bibitem [{\citenamefont {Yang}\ \emph {et~al.}(2018)\citenamefont {Yang}, \citenamefont {Culpitt},\ and\ \citenamefont {Hammes-Schiffer}}]{Yang2018}%
  \BibitemOpen
  \bibfield  {author} {\bibinfo {author} {\bibfnamefont {Y.}~\bibnamefont {Yang}}, \bibinfo {author} {\bibfnamefont {T.}~\bibnamefont {Culpitt}},\ and\ \bibinfo {author} {\bibfnamefont {S.}~\bibnamefont {Hammes-Schiffer}},\ }\bibfield  {title} {\bibinfo {title} {{Multicomponent Time-Dependent Density Functional Theory: Proton and Electron Excitation Energies}},\ }\href {https://doi.org/10.1021/ACS.JPCLETT.8B00547/SUPPL_FILE/JZ8B00547_SI_001.PDF} {\bibfield  {journal} {\bibinfo  {journal} {J. Phys. Chem. Lett.}\ }\textbf {\bibinfo {volume} {9}},\ \bibinfo {pages} {1765} (\bibinfo {year} {2018})}\BibitemShut {NoStop}%
\bibitem [{\citenamefont {Skolnick}\ \emph {et~al.}(1998)\citenamefont {Skolnick}, \citenamefont {Fisher},\ and\ \citenamefont {Whittaker}}]{Skolnick1998}%
  \BibitemOpen
  \bibfield  {author} {\bibinfo {author} {\bibfnamefont {M.~S.}\ \bibnamefont {Skolnick}}, \bibinfo {author} {\bibfnamefont {T.~A.}\ \bibnamefont {Fisher}},\ and\ \bibinfo {author} {\bibfnamefont {D.~M.}\ \bibnamefont {Whittaker}},\ }\bibfield  {title} {\bibinfo {title} {{Strong Coupling Phenomena in Quantum Microcavity Structures}},\ }\href {https://doi.org/10.1088/0268-1242/13/7/003} {\bibfield  {journal} {\bibinfo  {journal} {Semicond. Sci. Technol.}\ }\textbf {\bibinfo {volume} {13}},\ \bibinfo {pages} {645} (\bibinfo {year} {1998})}\BibitemShut {NoStop}%
\bibitem [{\citenamefont {Inoue}\ \emph {et~al.}(2013)\citenamefont {Inoue}, \citenamefont {Asano}, \citenamefont {{De Zoysa}}, \citenamefont {Oskooi},\ and\ \citenamefont {Noda}}]{Inoue2013}%
  \BibitemOpen
  \bibfield  {author} {\bibinfo {author} {\bibfnamefont {T.}~\bibnamefont {Inoue}}, \bibinfo {author} {\bibfnamefont {T.}~\bibnamefont {Asano}}, \bibinfo {author} {\bibfnamefont {M.}~\bibnamefont {{De Zoysa}}}, \bibinfo {author} {\bibfnamefont {A.}~\bibnamefont {Oskooi}},\ and\ \bibinfo {author} {\bibfnamefont {S.}~\bibnamefont {Noda}},\ }\bibfield  {title} {\bibinfo {title} {{Design of Single-Mode Narrow-Bandwidth Thermal Emitters for Enhanced Infrared Light Sources}},\ }\href {https://doi.org/10.1364/JOSAB.30.000165} {\bibfield  {journal} {\bibinfo  {journal} {J. Opt. Soc. Am. B}\ }\textbf {\bibinfo {volume} {30}},\ \bibinfo {pages} {165} (\bibinfo {year} {2013})}\BibitemShut {NoStop}%
\bibitem [{\citenamefont {{De Santis}}\ \emph {et~al.}(2020)\citenamefont {{De Santis}}, \citenamefont {Storchi}, \citenamefont {Belpassi}, \citenamefont {Quiney},\ and\ \citenamefont {Tarantelli}}]{DeSantis2020}%
  \BibitemOpen
  \bibfield  {author} {\bibinfo {author} {\bibfnamefont {M.}~\bibnamefont {{De Santis}}}, \bibinfo {author} {\bibfnamefont {L.}~\bibnamefont {Storchi}}, \bibinfo {author} {\bibfnamefont {L.}~\bibnamefont {Belpassi}}, \bibinfo {author} {\bibfnamefont {H.~M.}\ \bibnamefont {Quiney}},\ and\ \bibinfo {author} {\bibfnamefont {F.}~\bibnamefont {Tarantelli}},\ }\bibfield  {title} {\bibinfo {title} {{PyBERTHART: A Relativistic Real-Time Four-Component TDDFT Implementation Using Prototyping Techniques Based on Python.}},\ }\href {https://doi.org/10.1021/acs.jctc.0c00053} {\bibfield  {journal} {\bibinfo  {journal} {J. Chem. Theory Comput.}\ }\textbf {\bibinfo {volume} {16}},\ \bibinfo {pages} {2410} (\bibinfo {year} {2020})}\BibitemShut {NoStop}%
\bibitem [{\citenamefont {Habershon}\ and\ \citenamefont {Manolopoulos}(2009)}]{Habershon2009ZPE}%
  \BibitemOpen
  \bibfield  {author} {\bibinfo {author} {\bibfnamefont {S.}~\bibnamefont {Habershon}}\ and\ \bibinfo {author} {\bibfnamefont {D.~E.}\ \bibnamefont {Manolopoulos}},\ }\bibfield  {title} {\bibinfo {title} {{Zero Point Energy Leakage in Condensed Phase Dynamics: An Assessment of Quantum Simulation Methods for Liquid Water}},\ }\href {https://doi.org/10.1063/1.3276109} {\bibfield  {journal} {\bibinfo  {journal} {J. Chem. Phys.}\ }\textbf {\bibinfo {volume} {131}},\ \bibinfo {pages} {244518} (\bibinfo {year} {2009})}\BibitemShut {NoStop}%
\bibitem [{Sim()}]{Simpetus}%
  \BibitemOpen
  \href@noop {} {\bibinfo {title} {{Simpetus Projects: MEEP Thermal Radiation}}},\ \bibinfo {howpublished} {\url{http://www.simpetus.com/projects.html}},\ \bibinfo {note} {accessed: 2025-12-05}\BibitemShut {NoStop}%
\end{thebibliography}

%

    \end{document}